\documentclass[aps,prb,preprint,preprintnumbers,eqsecnum,amsmath,amssymb,showpacs]{revtex4}
\usepackage{graphicx}
\graphicspath{{/data/rhaertle/PaperI/submit22-01-2008/}}

\newcommand{\ee}{\end{equation}}
\newcommand{\be}{\begin{equation}}
\newcommand{\bea}{\begin{eqnarray}}
\newcommand{\eea}{\end{eqnarray}}
\newcommand{\bml}{\begin{subequations}} 
\newcommand{\eml}{\end{subequations}}

\newcommand{\meanT}[1]{\left\langle \text{T}_{c}\hspace{0.5mm} #1 \right\rangle}
\newcommand{\mean}[1]{\left\langle #1 \right\rangle}
\newcommand{\im}[1]{\text{Im}\hspace{-0.51mm}\left[ #1 \right]}
\newcommand{\unit}[1]{\hspace{.5mm}\text{\small #1}}

\begin{document}

\title{Multimode vibrational effects in single molecule conductance: 
A nonequilibrium Green's function approach}

\author{R.\ H\"artle, C.\ Benesch, and M.\ Thoss}

\affiliation{Department of Chemistry,
  Technical University of Munich,
  Lichtenbergstr.\ 4, D-85747 Garching, Germany}

\date{\today}

\begin{abstract}
The role of multimode vibrational dynamics in electron transport 
through single molecule junctions is investigated. The study is based on a generic model, which 
describes charge transport through a single molecule that is attached to 
metal leads.
To address vibrationally-coupled electron transport, 
we employ a nonequilibrium Green's function approach
that extends a method recently proposed by Galperin et al.\ [Phys.\ Rev.\ B {\bf 73}, 045314 (2006)] to multiple vibrational modes.
The methodology is applied to  two systems: a generic model 
with two vibrational degrees of freedom and  
benzenedibutanethiolate covalently bound to gold electrodes.
The results show that the coupling to multiple vibrational modes can have a significant
effect on the conductance of a molecular junction.
In particular, we demonstrate the 
effect of electronically induced coupling between different vibrational modes 
and study  nonequilibrium vibrational effects by calculating the current-induced
excitation of vibrational modes. 
\end{abstract}


\maketitle

\section{Introduction}

Experimental studies of single molecule conductance\cite{Reed97,Reichert02,Chen07,Selzer06} 
have revealed a wealth of interesting transport phenomena
and have stimulated great interest in the basic mechanisms that determine electron transport 
on the molecular scale.\cite{Haenggi02,Nitzan03,Cuniberti05}
Thereby, effects due to coupling between electronic and nuclear degrees of freedom
have been of particular interest.\cite{Galperin07} 
The small size and the low mass of molecules may result in strong 
coupling between electronic and nuclear degrees of  freedom.\cite{Park00,Pasupathy05}
Vibrational structures in molecular conductance have been observed for a variety of different 
systems.\cite{Park00,Gaudioso01,Kushmerick04,Qiu04,Djukic05,Sapmaz06,Thijssen06,Boehler07,Parks07,Ogawa07}
Electronic-vibrational coupling can result in excitation of the vibrational modes 
of the molecular bridge as well as local heating. \cite{Sapmaz06}
Conformational changes of the geometry of the conducting molecule are 
possible mechanisms for switching behavior and negative differential resistance
\cite{Gaudioso00}.
Furthermore, the observation of vibrational structures in conductance measurements allows the unambiguous identification of the molecule in the junction.

These experimental findings have inspired great interest in the theoretical modeling and
simulation of vibrationally-coupled charge transport in molecular junctions.\cite{Emberly00,Schoeller01,Ness01,Troisi03,Pecchia04,Chen05,May06,Lehmann04,Koch05,Wegewijs05,Jiang06,Ryndyk06,Galperin06,Galperin07,Gagliardi07,Sergueev07,Frederiksen07,Toroker07}
A variety of different approaches have been used to study the influence of the 
vibrational degrees of freedom on single molecule conductance, including
inelastic scattering theory, density matrix approaches and 
nonequilibrium Green's function methods. 
Green's functions are particularly well-suited
to study many-body and vibrational nonequilibrium effects.
Employing perturbation theory, nonequilibrium Green's function methods have been applied
in the off-resonant tunneling regime to study, e.g., inelastic tunneling spectra.\cite{Pecchia04,Chen05,Frederiksen07}
In this regime, the effective electronic-vibrational coupling is
 typically small and perturbation theory valid.
Galperin et al.\ have recently proposed a method that
allows the study of vibrational effects in the resonant transport regime.\cite{Galperin06}
This method is based on a polaron transformation of the Hamiltonian and  
employs perturbation theory within a self-consistent scheme to solve the equations of motion
for the nonequilibrium Green's function. 

In the original formulation,\cite{Galperin06} the method of Galperin et al.\ is limited to the treatment of 
a single vibrational mode that is coupled to a thermal bath.
In this paper, we extend this Green's function method 
to allow the treatment of several vibrational degrees of freedom.
Moreover,  we outline a scheme that allows
the calculation of current-induced vibrational excitation and thus to study
vibrational nonequilibrium effects.
The methodology is applied to  two different systems: a generic model of
a molecular junction with two active vibrational modes as well as charge transport through 
benzenedibutanethiolate covalently bound to gold electrodes based on a 
recently developed\cite{note_Benesch08b} first-principles model.

\section{Theory}
\label{secondsection}

\subsection{Model}

To study vibrationally-coupled electron transport through a molecular junction,
we consider a generic tight-binding model, \cite{Cizek04,Galperin06}
where a single electronic state localized on the molecule 
 is coupled to respective states in the left (L) and right (R) 
leads by  tunneling matrix elements $V_{k}$,
\begin{eqnarray}\label{hamiltonian}
H &=&   \epsilon_{0}  c^{\dagger}c + \sum_{k\in\text{L,R}} \epsilon_{k} 
c^{\dagger}_{k}c_{k} +\sum_{k\in\text{L,R}} (  V_{k}c^{\dagger}_{k}c + 
V_{k}^{*}c^{\dagger}c_{k} )\\
&& + \sum_{\alpha} \Omega_{\alpha} a_{\alpha}^{\dagger}a_{\alpha} + 
\sum_{\alpha}   \lambda_{\alpha} Q_{\alpha} ( c^{\dagger}c - \delta ) + \sum_{\beta} 
\omega_{\beta} b_{\beta}^{\dagger}b_{\beta} + \sum_{\alpha\beta} Q_{\alpha} 
U_{\alpha\beta} Q_{\beta}. \nonumber
\end{eqnarray}
Here, $c_k^{\dagger}$ and $c^{\dagger}$ are operators that create an electron 
in the leads and on the molecular bridge, respectively, and $\epsilon_{k}$, $\epsilon_{0}$ 
denote the energies of the corresponding electronic states.
The nuclear degrees of freedom of the molecule are described in  Eq.\ (\ref{hamiltonian}) 
within the
harmonic approximation employing the normal modes of the neutral junction in equilibrium 
at zero bias. 
Thereby,  $a_{\alpha}^{\dagger}$ denotes the creation operator for
a normal mode of the neutral molecule  with frequency $\Omega_{\alpha}$
and $Q_{\alpha}=(a_{\alpha}+a_{\alpha}^{\dagger})$ is the corresponding displacement operator.

If an external bias is applied to the junction, electrons will be transferred from the 
leads to the junction and vice versa. Accordingly, the potential energy  for the nuclei
will be distorted.  The corresponding interaction potential is 
approximated employing a linear expansion  in the displacement operator 
$Q_{\alpha}$ around the equilibrium geometry of the
neutral junction, resulting in the electronic-vibrational interaction term 
$\sum_{\alpha}   \lambda_{\alpha} Q_{\alpha} ( c^{\dagger}c - \delta)$  in Eq.\ (\ref{hamiltonian}).
The charge density, to which this interaction 
potential is linked, depends on whether the molecular state through which the transport
takes place is unoccupied in equilibrium
($\delta=0$, corresponding, e.g., to charge transport through the LUMO) 
or occupied ($\delta=1$, corresponding, e.g., to the HOMO).\cite{note1}
The corresponding shift in the equilibrium geometry of mode $\alpha$
 is given by $\bigtriangleup Q_{\alpha}=2 \lambda_{\alpha}/\Omega_{\alpha}$.

To account for the effect of relaxation of the vibrational modes, 
induced by anharmonic interactions
and coupling to phonons in the electrodes,
we adopt a linear response model for vibrational relaxation.\cite{Cizek04,Galperin06}
Within this model, the normal modes of the molecule are 
coupled linearly in Eq.\ (\ref{hamiltonian})
to a thermal bath of secondary modes. Thereby, $b_{\beta}^{\dagger}$ denotes 
the creation operator
for a bath mode with frequency $\omega_{\beta}$ and $U_{\alpha\beta}$ determines the
system-bath coupling strength. 
All properties of the bath, which influence the dynamics of the system
are characterized by the spectral density\cite{Weiss99}
\begin{eqnarray}\label{spec_dens}
J_{\alpha}(\omega)&=& \sum_{\beta}\vert 
U_{\alpha\beta}\vert^{2}\delta(\omega-\omega_{\beta}).
\end{eqnarray}
In the applications considered below, we have used an Ohmic spectral density 
\begin{eqnarray}\label{Ohmic}
J_{\alpha}(\omega)&=& \eta_{\alpha}\omega
e^{-\omega/\omega_{c\alpha}}.
\end{eqnarray}
Here, the characteristic frequency $\omega_{c\alpha}$  defines the maximum of the 
spectral density and the overall coupling strength is determined by $\eta_{\alpha}$. 
Both values may depend on the specific
system mode $\alpha$.

To apply (self-consistent) perturbation theory within the nonequilibrium Green's functions 
approach considered below, it is expedient
to remove the direct coupling terms between electrons and 
vibrations in the Hamiltonian, $\lambda_{\alpha}Q_{\alpha}(c^{\dagger}c-\delta)$. To this end, we 
apply a 
standard canonical transformation, also referred to as Lang-Firsov 
(or small polaron) transformation\cite{Mahan81,Lang63}
\begin{eqnarray}\label{trans_ham}
\overline{H} &\equiv& \text{e}^{S} H \text{e}^{-S},  \\
&=&   \overline{\epsilon}_{0}  c^{\dagger}c + \sum_{k\in\text{L,R}} 
\epsilon_{k} c^{\dagger}_{k}c_{k} +\sum_{k\in\text{L,R}} (  
V_{k} X c^{\dagger}_{k}c + 
V_{k}^{*}X^{\dagger}c^{\dagger}c_{k} ) \nonumber\\
&& + 
\textbf{A}^{\dagger}\textbf{W}_{\text{a}}\textbf{A} + \textbf{B}^{\dagger} \textbf{W}_{\text{b}} \textbf{B}+\textbf{Q}_{\text{a}}^{\dagger}\textbf{U}\textbf{Q}
_{\text{b}}, \nonumber
\end{eqnarray}
with $X = 
\text{exp}\left(i\textbf{P}_{\text{a}}^{\dagger}\textbf{W}_{\text{a}}^{-1}\mathbf{\Lambda}
\right)$ and $S=-i 
\textbf{P}_{\text{a}}^{\dagger}\textbf{W}_{\text{a}}^{-1}\mathbf{\Lambda} 
( c^{\dagger}c - \delta)$. For notational convenience we introduce the vectors 
$\left(\textbf{Q}_{\text{a}}\right)_{\alpha}=Q_{\alpha}$, 
$\left(\textbf{P}_{\text{a}}\right)_{\alpha}=-i(a_{\alpha}-a_{\alpha}^{\dagger})
$,
$\left(\textbf{A}\right)_{\alpha}=a_{\alpha}$, 
$\left(\mathbf{\Lambda}\right)_{\alpha}=\lambda_{\alpha}$, the matrices  
$\left(\textbf{W}_{\text{a}}\right)_{\alpha\alpha'}=\delta_{\alpha\alpha'}\Omega
_{\alpha}$, $\left(\textbf{U}\right)_{\alpha\beta}=U_{\alpha\beta}$, as well as
respective quantities for the bath modes. 
It is noted that, in the presence of a bath, the decoupling requires 
that the eigenvalues of the matrix 
$\left(4\textbf{U}\textbf{W}_{\text{b}}^{-1}
\textbf{U}^{\dagger}\textbf{W}_{\text{a}}^{-1}\right)$ are smaller than 
unity,\cite{Galperin06} corresponding to weak system-bath coupling.

The electronic-vibrational coupling manifests itself in 
the transformed Hamiltonian (\ref{trans_ham}) in a polaron shift of the electronic energy
\begin{eqnarray} \label{polaron_shift}
\epsilon_{0}\rightarrow\overline{\epsilon}_{0}&=&
\begin{cases}
\epsilon_{0}-\mathbf{\Lambda}^{\dagger}\textbf{W}_{\text{a}}^{-1}\mathbf{\Lambda},&
\text{$\delta=0$},\\
\epsilon_{0}+\mathbf{\Lambda}^{\dagger}\textbf{W}_{\text{a}}^{-1}\mathbf{\Lambda},& \text{$\delta=1$},\\
\end{cases} 
\end{eqnarray}
and in the shift generator 
$X$ that dresses the molecule-lead couplings $V_{k}$.

\subsection{Nonequilibrium Green's function approach}

To describe transport properties for the model introduced above, we employ
a nonequilibrium Green's function method, which was proposed by Galperin et al.\cite{Galperin06}
Here, we generalize this method for applications to multiple vibrational system modes and 
outline how this method can be used to calculate vibrational properties.

The central quantity in nonequilibrium Green's function theory is the
electronic Green's function on the molecular bridge
\begin{eqnarray}
G(\tau,\tau') &=&-i \meanT{c(\tau)c^{\dagger}(\tau')}_{H},\\
&=& -i 
\meanT{c(\tau)c^{\dagger}(\tau')X(\tau)X^{\dagger}(\tau')}
_{\overline{H}},\nonumber
\end{eqnarray}
where $\text{T}_{c}$ denotes the time-ordering operator along the Keldysh 
contour\cite{Keldysh65} and the subscripts $H$/$\overline{H}$ indicate 
the Hamiltonian that is used in the calculation of the respective expectation value.
Most
expectation values considered below refer to $\overline{H}$, for which the 
corresponding subscript is omitted in the following.

Following Galperin et al.,\cite{Galperin06} the electronic Green's function $G(\tau,\tau')$ is factorized,
\begin{eqnarray}
G(\tau,\tau') &\approx & G_{c}(\tau,\tau') 
\meanT{X(\tau)X^{\dagger}(\tau')},
\end{eqnarray}
 into a correlation function of the shift generator, 
$\meanT{X(\tau)X^{\dagger}(\tau')}$, and an electronic 
Green's function, $G_{c}(\tau,\tau')$,
with
\begin{eqnarray}
G_{c}(\tau,\tau')= -i  \meanT{c(\tau)c^{\dagger}(\tau')}_{\overline{H}}.
\end{eqnarray}
This factorization is valid in the limit of weak molecule-lead coupling corresponding
to a relatively long residence time of the electron on the molecular bridge. This
is the regime, where vibrational effects are expected to be particularly 
pronounced.\cite{note_Benesch08b}

Based on the Hamiltonian $\overline{H}$ and employing the equation of motion  for the 
electronic Green's function, the following equation for
$G_{c}(\tau,\tau')$ is obtained\cite{Galperin06,Haug98}
\begin{eqnarray}
G_{c}(\tau,\tau') &=& G_{c}^{0}(\tau,\tau') + 
\int\hspace{-1mm}\text{d}\tau_{1}\text{d}\tau_{2}\hspace{1mm} 
G_{c}^{0}(\tau,\tau_{1}) \Sigma_{c}(\tau_{1},\tau_{2}) 
G_{c}^{0}(\tau_{2},\tau'),\label{timedepDyKe}
\end{eqnarray}
where  $G_{c}^{0}$ 
denotes the electronic Green's function for vanishing electronic coupling (i.e. 
$V_{k} =0$).
The self energy $\Sigma_{c}(\tau,\tau')$, introduced in Eq.\ (\ref{timedepDyKe}), 
comprises  all interactions of the 
electronic degrees of freedom on the molecule with those in the leads 
and the vibrational modes and is given by
\begin{eqnarray}
\label{schematicdressedselfenergy}
\Sigma_{c}(\tau,\tau') &=& \sum_{k\in\text{L,R}} \left\vert V_{k}\right\vert^{2} 
g_{k}(\tau,\tau')\meanT{X(\tau')X^{\dagger}(\tau)},\\
&\equiv& \Sigma_{c}^{0}(\tau,\tau') 
\meanT{X(\tau')X^{\dagger}(\tau)}, \nonumber 
\end{eqnarray}
with $g_{k}=-i\langle \text{T}_{c}\hspace{0.5mm} c_{k}(\tau)c_{k}^{\dagger}(\tau')\rangle$. The above expression holds to second order in $V_{k}$, i.e.\ for weak molecule-lead coupling. 

To extend the validity to moderate coupling, which is particularly important to describe
vibrational effects in resonant electron transport, higher order terms are taken 
into account. To this end, a self-consistent scheme is introduced by replacing
the last $G_{c}^{0}$ in the integral kernel of Eq.\ (\ref{timedepDyKe}) by $G_{c}$,\cite{Haug98,Galperin06}
\begin{eqnarray}\label{timedepDyKesc}
G_{c}(\tau,\tau') &=& G_{c}^{0}(\tau,\tau') + 
\int\hspace{-1mm}\text{d}\tau_{1}\text{d}\tau_{2}\,
G_{c}^{0}(\tau,\tau_{1}) \Sigma_{c}(\tau_{1},\tau_{2}) 
G_{c}(\tau_{2},\tau'). 
\end{eqnarray}
It is noted that Eq.\ (\ref{timedepDyKesc})  gives the exact electronic Green's function
for vanishing electronic-vibrational coupling (i.e.\ $\lambda_{\alpha}=0$).

Projection of Eq.\ (\ref{timedepDyKesc}) onto the real time axis, according to 
analytic continuation rules,\cite{Langreth76} and Fourier transformation of the resulting equations 
gives a Dyson equation
\bml \label{GCDysonKeldysh}
\begin{eqnarray}
G_{c}^{\text{r}}(E) &=& G_{c}^{0,\text{r}} + G_{c}^{0,\text{r}}(E) 
\Sigma_{c}^{\text{r}}(E) G_{c}^{\text{r}}(E),
\end{eqnarray}
and a Keldysh equation
\begin{eqnarray}
G_{c}^{<}(E) &=& G_{c}^{\text{r}}(E) \Sigma_{c}^{<}(E) G_{c}^{\text{a}}(E),
\end{eqnarray}
\eml
where we have introduced the retarded ($G_{c}^{\text{r}}$), advanced
($G_{c}^{\text{a}}$), and lesser ($G_{c}^{<}$) Green's functions and the corresponding self 
energies.
It should be emphasized that the derivation of the compact form of these equations
requires time-translational invariance and the existence of a 
steady-state transport regime.

So far, we have only considered the electronic part of the Green's
function $G(\tau,\tau')$. The solution of the equations requires
also the calculation of the shift 
generator correlation function $\meanT{X(\tau)X^{\dagger}(\tau')}$. 
Using a cumulant expansion up to second order in the 
vibronic coupling parameters $\lambda_{\alpha}/\Omega_{\alpha}$ yields\cite{Galperin06,Mahan81}
\begin{eqnarray}
\label{shiftgenandcorrmatr}
\meanT{X(\tau)X^{\dagger}(\tau')} &=& \text{exp}\left[i 
\mathbf{\Lambda}^{\dagger}\textbf{W}_{\text{a}}^{-1} 
\left(\textbf{D}(\tau,\tau')-\textbf{D}(\tau,\tau)\right) \textbf{W}_{\text{a}}^{-1}\mathbf{\Lambda}\right],
\end{eqnarray}
where 
$\textbf{D}(\tau,\tau')=-i\meanT{\textbf{P}_{\text{a}}(\tau)\textbf{P}_{\text{a}
}^{\dagger}(\tau')}$ denotes the correlation matrix of the momentum vector 
$\textbf{P}_{\text{a}}$, in the following referred to as the vibrational Green's function. In the limit $V_{k}\rightarrow0$, Eq.\ (\ref{shiftgenandcorrmatr}) constitutes an analytically exact expression. Employing again an equation of motion  approach, one can derive 
Dyson/Keldysh equations for  the vibrational Green's function $\textbf{D}(\tau,\tau')$.
 The resulting self energy term 
$\mathbf{\Pi}_{\text{el}}(\tau,\tau')$ describes interactions between 
electrons and vibrations.
This self energy was obtained by Galperin et al.\ for a single 
vibrational mode.\cite{Galperin06} Employing  the same level of approximations, we obtain  
for multiple vibrational modes
\begin{eqnarray}
\mathbf{\Pi}_{\text{el}}(\tau,\tau') & = & - i  
\textbf{W}_{\text{a}}^{-1} \mathbf{\Lambda} 
\mathbf{\Lambda}^{\dagger} \textbf{W}_{\text{a}}^{-1} \left( 
\Sigma_{c}(\tau,\tau') G_{c}(\tau',\tau) + \Sigma_{c}(\tau',\tau) 
G_{c}(\tau,\tau')  \right), \label{SigmaintoPiel}
\end{eqnarray}
and
\bml \label{DDysKeld}
\begin{eqnarray}
\textbf{D}^{\text{r}}(E) &=& \textbf{D}^{0,\text{r}} + 
\textbf{D}^{0,\text{r}}(E) \mathbf{\Pi}_{\text{el}}^{\text{r}}(E) 
\textbf{D}^{\text{r}}(E),\\
\textbf{D}^{<}(E) &=& 
\textbf{D}^{\text{r}}(E) \mathbf{\Pi}_{\text{el}}^{<}(E) 
\textbf{D}^{\text{a}}(E).
\end{eqnarray}
\eml

The off-diagonal elements of the self energy matrix $\mathbf{\Pi}_{\text{el}}$
describe
interactions between different vibrational modes mediated by the electronic degrees of freedom. 
The matrix $\textbf{D}^{0}$  describes momentum correlations of the vibrations in 
thermal equilibrium and plays the same role for the vibrational Green's function
$\textbf{D}$ as $G_{c}^{0}$ for the electronic Green's function $G_{c}$. 
In the presence of a thermal bath, the matrix $\textbf{D}^{0}$
need not to be of diagonal form, especially 
if the bath degrees of freedom couple the various vibrational modes with each 
other.

The equations for the electronic and vibrational Green's functions outlined above
have to be solved self-consistently. We employ the following self-consistent
scheme:\cite{Galperin06}
As a starting point we use the free electronic and vibrational Green's functions.
The free vibrational Green's function $\textbf{D}^{0}$
enters the shift generator correlation function
according to Eq.\ (\ref{shiftgenandcorrmatr}).
Next, the shift generator correlation function is convoluted  with the bare self energy 
$\Sigma_{c}^{0}$, which gives the dressed self energy $\Sigma_{c}$ according to 
Eq.\ (\ref{schematicdressedselfenergy}).
The dressed self energy $\Sigma_{c}$, which now contains 
interactions of the electronic degrees of freedom on the molecule with both the leads and the vibrations, is 
inserted into the electronic Green's function $G_{c}$, Eqs.\ (\ref{GCDysonKeldysh}), 
and into the self energy term  $\mathbf{\Pi}_{\text{el}}$ of the  
vibrational Green's function, Eq.\ (\ref{SigmaintoPiel}). The new Green's function
\textbf{D} is obtained from the Dyson/Keldysh equations (\ref{DDysKeld}), 
and enters the shift generator correlation function.  Thus, the self-consistent cycle
closes, and all steps can be repeated with the updated 
shift generator correlation function.
Convergence is reached as soon as the variation of the electronic occupation number, 
$n_{c}=\im{G_{c}^{<}(t=0)}$, between 
subsequent iteration steps falls below a threshold of $10^{-7}$. This way, we 
obtain self-consistent solutions of Eqs.\ (\ref{GCDysonKeldysh}) and 
(\ref{DDysKeld}).

\subsection{Observables of interest}

Several observables can be considered to study the effects of vibrational motion on charge
transport through single molecule junctions.
Here, we will focus on the current-voltage characteristic, the differential conductance and the vibrational
nonequilibrium distribution.

Based on the self-consistent result for the Green's function, the 
current through lead 
$\text{K}$ ($\text{K}\in\lbrace\text{L},\text{R}\rbrace$) induced by an external dc-bias $\Phi$
is obtained employing the formula of
Meir, Wingreen and 
Jauho\cite{Meir92,Jauho94,Haug98}
\begin{eqnarray}
\label{current}
I_{\text{K}} = \frac{2e}{\hbar} \int \frac{\text{d}E}{2\pi} \hspace{1mm} \left[ 
\Sigma_{c,\text{K}}^{<}(E)G_{c}^{>}(E)  - \Sigma_{c,\text{K}}^{>}(E)G_{c}^{<}(E) 
\right],
\end{eqnarray}
with
\begin{eqnarray}
\Sigma_{c,\text{K}}(\tau,\tau') &=&  \sum_{k\in\text{K}} \left\vert V_{k}\right\vert^{2} 
g_{k}(\tau,\tau') 
\meanT{X(\tau')X^{\dagger}(\tau)},\\
&\equiv& \Sigma_{c,\text{K}}^{0}(\tau,\tau') 
\meanT{X(\tau')X^{\dagger}(\tau)}. \nonumber 
\end{eqnarray}
Thereby, the factor two accounts for spin degeneracy.
The differential conductance is given by $\text{g}=\text{d}I/\text{d}\Phi$.

It is noted that the  scheme outlined above conserves the number of  electrons and thus  obeys 
Kirchhoff's law, $I_{\text{L}}=-I_{\text{R}}$. Furthermore, as there is no direct 
electron-vibrational coupling term in the Hamiltonian $\overline{H}$, the symmetrized current $I=1/2\left(I_{\text{L}}-I_{\text{R}}\right)$
can be expressed in terms of a transmission function $\mathcal{T}(E)$
\begin{eqnarray}
I &=& \frac{2e}{\hbar}\int 
\frac{\text{d}E}{2\pi} \left(f_{\text{L}}(E)-f_{\text{R}}(E)\right) 
\mathcal{T}(E),
\end{eqnarray}
with
\begin{eqnarray}
\mathcal{T}(E)&=&\frac{\Gamma_{\text{L}}(E)\Gamma_{\text{R}}(E)}{\Gamma_{\text{L}}(E)+
\Gamma_{\text{R}}(E)}i\left( G_{c}^{\text{r}}(E)-G_{c}^{\text{a}}(E)\right).
\end{eqnarray}
Here, we have introduced the nonequilibrium distribution function
\begin{eqnarray}
f_{\text{K}}(E) &=& 
\frac{\im{\Sigma_{c,\text{K}}^{<}(E)}}{\Gamma_{\text{K}}(E)},
\end{eqnarray}
and the width function
\begin{eqnarray}
\Gamma_{\text{K}}(E)&=&-2\im{\Sigma_{c,\text{K}}^{\text{r}}(E)}.
\end{eqnarray}

Another interesting observable to investigate vibrationally-coupled electron transport in 
molecular junctions is the nonequilibrium vibrational distribution. 
Due to current-induced excitation and deexcitation of the vibrational modes,
the vibrational distribution in the stationary state
may differ significantly from its equilibrium distribution.
To study such vibrational nonequilibrium effects,  we consider in this work 
the average occupation number of different vibrational modes
\begin{eqnarray}
n_{\alpha} &=& \mean{a^{\dagger}_{\alpha} a_{\alpha}}_{H}.
\end{eqnarray} 
Within the self-consistent
Green's function approach employed here, $n_{\alpha}$ is given (up 
to second order in the system-bath interaction $U_{\alpha\beta}$) by the expression
\begin{eqnarray}
\label{nBocc}
n_{\alpha} &=& - \left( A_{\alpha} + \frac{1}{2} \right)  
\im{\left(\textbf{D}(t=0)\right)_{\alpha\alpha}} - 
\left(B_{\alpha}+\frac{1}{2}\right) + 
\frac{\lambda_{\alpha}^{2}}{\Omega_{\alpha}^{2}} \begin{cases}
n_{c}, & \delta=0, \\
1- n_{c}, & \delta=1, \\
\end{cases} 
\end{eqnarray}
with
\bml
\begin{eqnarray}
A_{\alpha}&=&\sum_{\beta} 
\frac{U_{\alpha\beta}^{2}\omega_{\beta}}{\Omega_{\alpha}(\omega_{\beta}^{2}-
\Omega_{\alpha}^{2})},\\
 B_{\alpha}&=&\sum_{\beta} 
\frac{U_{\alpha\beta}^{2}}{(\omega_{\beta}^{2}-\Omega_{\alpha}^{2})} 
(1+2N_{\text{B}}(\omega_{\beta})).
\end{eqnarray}
\eml
Thereby, $n_{c}= \mean{c^{\dagger}c}_{H}$ denotes the stationary population of the molecular electronic state. 
The derivation of Eq.\ (\ref{nBocc})
is outlined in the Appendix.

\section{Results and Discussion}

The methodology outlined above has been applied to two different systems:
A generic model for a molecular junction including two vibrational modes and charge transport
through benzenedibutanethiolate coupled to gold electrodes. In the latter system the 
four most strongly coupled  vibrational modes have been taken into account.

\subsection{Generic model system with two vibrational degrees of freedom}
\label{twovibrons}

First, we consider a generic model system with a single molecular 
orbital coupled to two vibrational degrees of freedom. The energy of the molecular state 
is  chosen as $\epsilon_{0}=1\unit{eV}$ and is located well above the Fermi energy of the 
leads, $\epsilon_{\text{F}}=0\unit{eV}$. Thus in equilibrium, the
molecular state
is  unoccupied (corresponding to $\delta=0$). 
The leads are modelled by a one-dimensional tight-binding model with
nearest-neighbor coupling constant $\beta = 2\unit{eV}$ and molecule-lead coupling strength $\nu=0.1\unit{eV}$.
This results in an unstructured  semi-elliptic conduction band with 
self energies\cite{Cizek04}
\bml
\begin{eqnarray}
\label{bareselfenergy}
\Sigma_{c,\text{K}}^{0,\text{r}}(E) &=& \Delta_{\text{K}}^{{0}}(E) - 
\frac{i}{2} \Gamma_{\text{K}}^{0}(E),\\
\Sigma_{c,\text{K}}^{0,<}(E)&=&if_{\text{K}}^{0}(E)\Gamma_{\text{K}}^{0}(E), \\
\Sigma_{c,\text{K}}^{0,>}(E)&=&-i(1-f_{\text{K}}^{0})(E)\Gamma_{\text{K}}^{0}(E),
\end{eqnarray}
\eml
where the corresponding level-width function reads
\begin{eqnarray}
\Gamma_{\text{K}}^{0}(E) &=& 
\frac{\nu^{2}}{\beta^{2}} \im{\sqrt{(E-\mu_{\text{K}})^{2}-4\beta^{2}}},  
\end{eqnarray}
and $\Delta_{\text{K}}^{0}$ is the Hilbert transform of $\Gamma_{\text{K}}^{0}$. 
Furthermore, $f_{\text{K}}^{0}$ denotes the Fermi distribution in the leads,
\begin{eqnarray}
f_{\text{K}}^{0}(E)&=&\frac{1}{1+\text{exp}\left(\frac{E-\mu_{\text{K}}}{k_{\text{B}}
T}\right)},
\end{eqnarray}
$\mu_{\text{L(R)}}$ is the chemical potential in the leads,
and $\Phi$ the bias voltage.
We assume the bias voltage to drop 
symmetrically at the right and the left contact, 
$\mu_{\text{L(R)}}= \epsilon_{\text{F}}\pm\Phi/2$.
In the results presented below, we have used a temperature of 
$k_{\text{B}}T = 1\unit{meV}$. This temperature is 
low enough to study vibrational features undistorted by thermal fluctuations ($k_{\text{B}}T\ll\Omega_{1(2)}$).

The parameters of the two vibrational modes of the model are given in 
Tab.\ \ref{partwovib}.
\begin{table}
\begin{tabular}{|ccc|ccc|ccc|}
\hline
& frequency &&& vibronic coupling &&& system-bath coupling & \\
\hline
&$\Omega_{1} = 0.10\unit{eV}$& & &$\lambda_{1}=0.06\unit{eV}$& & &$\eta_{1}=0.001$& 
\\
&$\Omega_{2} = 0.25\unit{eV}$& & &$\lambda_{2}=0.15\unit{eV}$& & &$\eta_{2}=0.001$& 
\\
\hline
\end{tabular}
\caption{\label{partwovib} Vibrational parameters of the model system with two vibrational modes.}
\end{table}
Each mode is coupled to  an Ohmic bath as described by Eqs.\ (\ref{spec_dens}),  (\ref{Ohmic}).
Thereby, the characteristic frequencies of the bath spectral density were chosen to coincide
with the frequency of the respective system mode, i.e.\ $\omega_{c\alpha} = \Omega_{\alpha}$.
A relatively weak system-bath coupling strength was used, $\eta_{\alpha} = 0.001$, corresponding
to vibrational relaxation times of about $0.1\text{--}1\unit{ps}$.
 In principle, the thermal bath couples the two 
vibrational modes with each other.  This interaction is neglected in the calculation 
presented below. Hence, the retarded projection of the 
correlation matrix $\textbf{D}^{0}$ has a diagonal form
\begin{eqnarray}
\textbf{D}^{0,\text{r}}_{\alpha\alpha'}(\omega) &=& \delta_{\alpha\alpha'} 
\frac{2\Omega_{\alpha}}{\omega^{2}-\Omega_{\alpha}^{2}-2\Omega_{\alpha}
\Pi^{\text{r}}_{\text{bath},\alpha}(\omega)},
\end{eqnarray}
with
\begin{eqnarray}
-2\im{\Pi^{\text{r}}_{\text{bath},\alpha}(\omega) }&=&2\pi J_{\alpha}(\omega). 
\end{eqnarray}

\begin{figure}
\resizebox{0.85\textwidth}{0.56\textwidth}{
\includegraphics{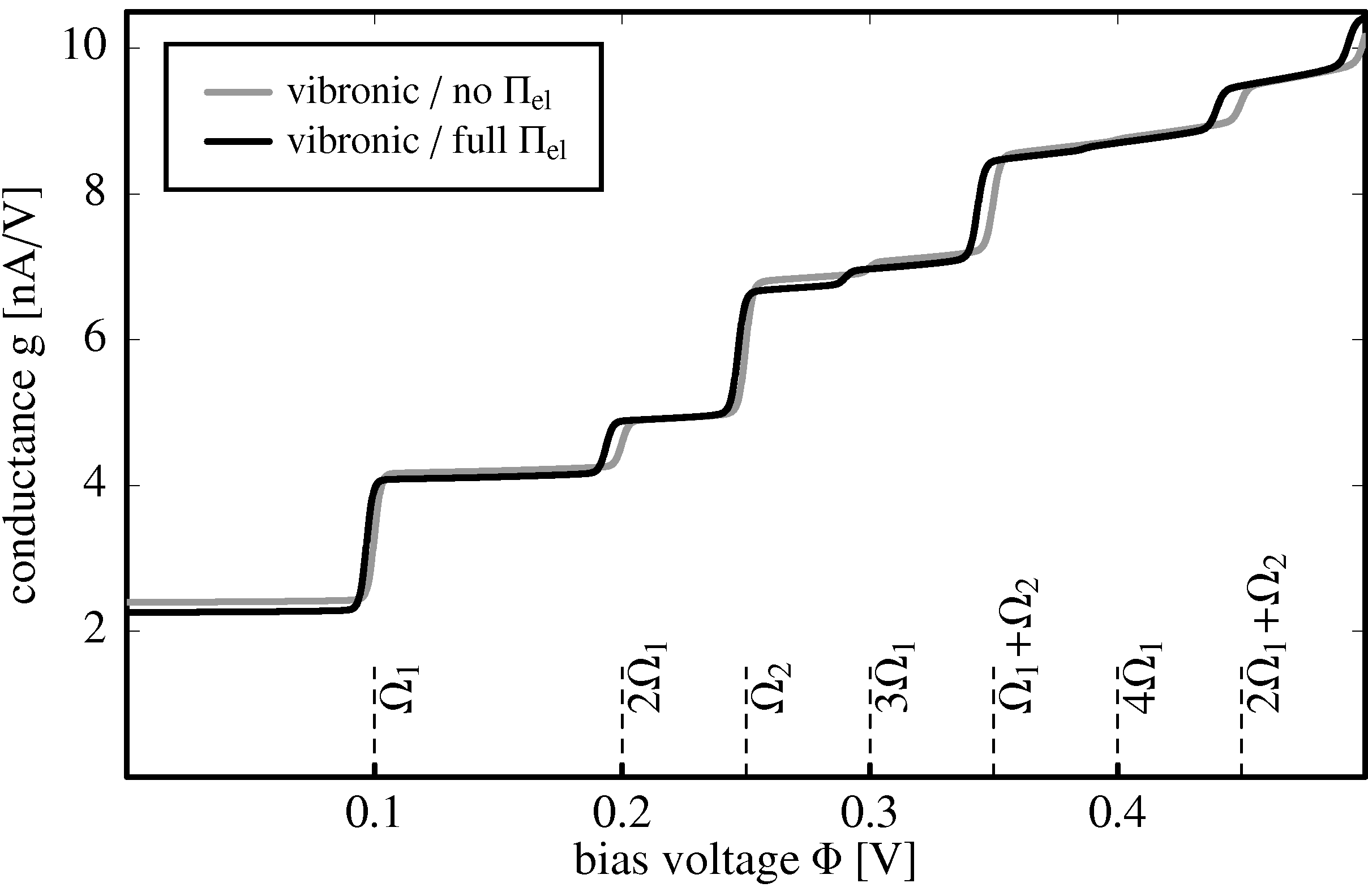}
}\\
\caption{\label{PaperI-IETS-twophon} Conductance g for the model with two vibrational modes
in the inelastic electron 
tunneling regime. The solid gray line refers to a calculation with vibrations 
in thermal equilibrium, $\mathbf{\Pi}_{\text{el}}=0$, while the solid black 
line presents results including nonequilibrium effects, 
$\mathbf{\Pi}_{\text{el}}\neq0$. The thin dashed lines indicate the voltages corresponding
to the opening of vibrationally inelastic channels.
}
\end{figure}

The  approach presented above can be used to describe two different 
regimes of electron transport: 
inelastic electron tunneling in the off-resonant regime and 
inelastic resonant electron transport.  
 Fig.\ \ref{PaperI-IETS-twophon} shows the  
conductance as a function of bias voltage in the inelastic electron tunneling regime,
$\Phi\lesssim1\unit{V}$.
In this off-resonant transport regime, electronic-vibrational coupling manifests itself in
steps of the conductance at voltages $\Phi=\Omega_{1},2\Omega_{1},\Omega_{2},..$ 
that correspond to the opening of 
inelastic channels.
These channels are related to the excitation of vibrational quanta
and will be referred to as emission channels in the following. 
The relative step heights reflect the 
respective transition probabilities determined by the corresponding
Franck-Condon factors. Absorptive 
channels, corresponding to deexcitation of vibrational quanta, 
play only a minor role in this regime. This is due to the low 
temperature and the small electric current, which is insufficient to drive the 
vibrational system far from equilibrium. 
The comparison of the results obtained 
with (black line) and without (gray line) 
vibrational self energy $\mathbf{\Pi}_{\text{el}}$
shows that in the off-resonant regime vibrational nonequilibrium effects
manifest themselves in a shift of the  conductance steps.

\begin{figure}
\resizebox{0.85\textwidth}{0.56\textwidth}{
\includegraphics{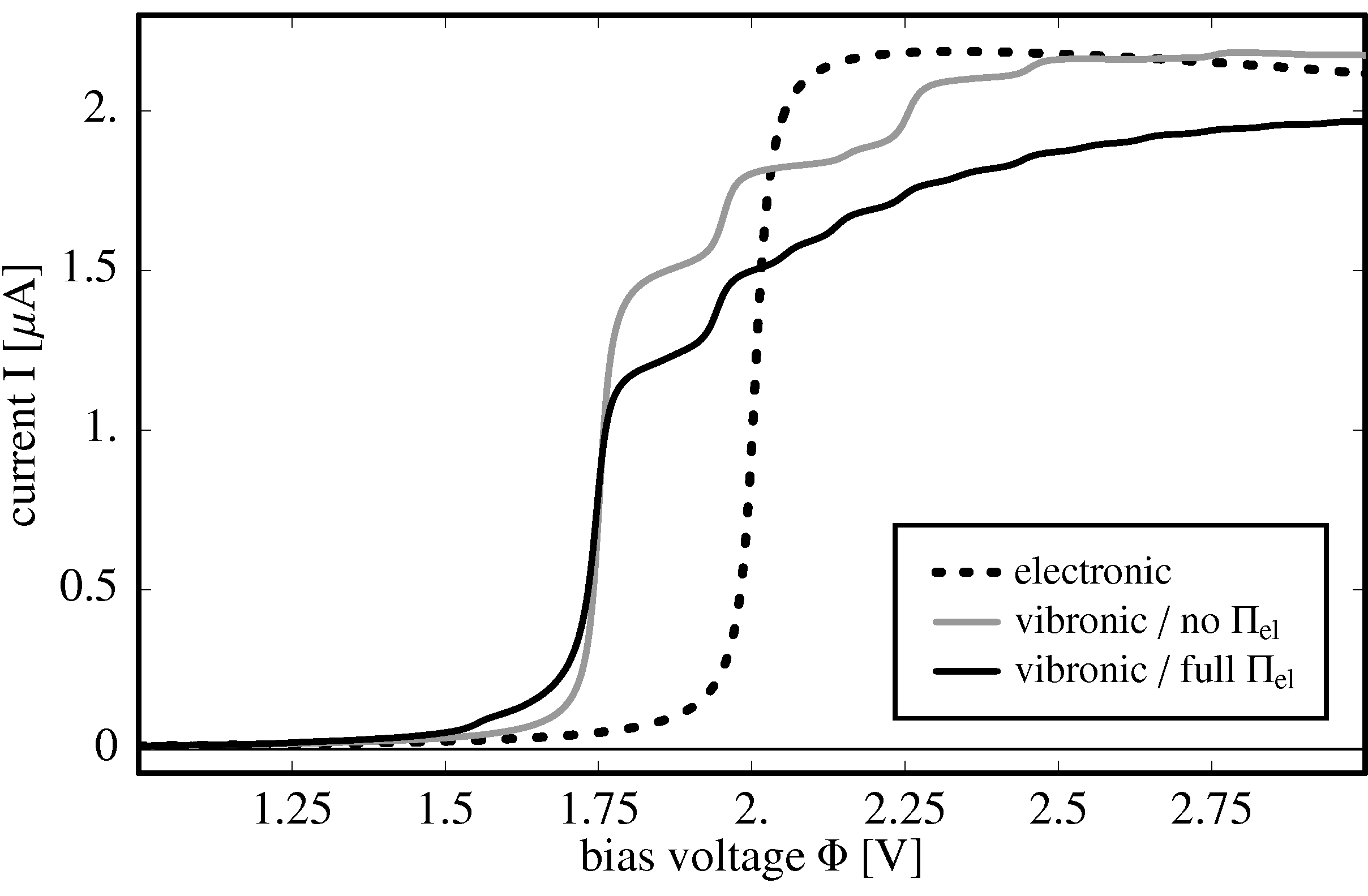}
}\\
\resizebox{0.85\textwidth}{0.56\textwidth}{
\includegraphics{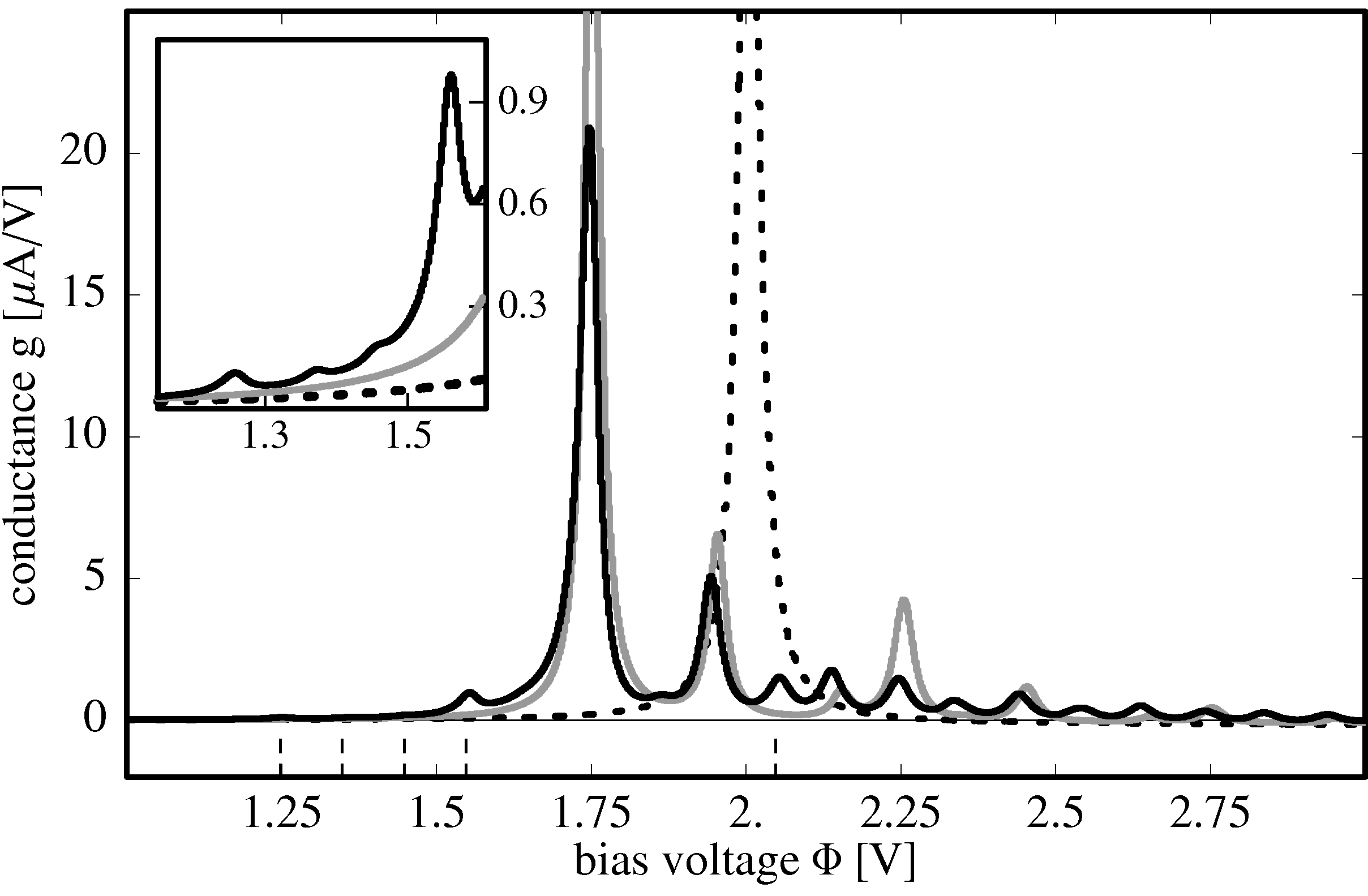}
}\\
\caption{\label{PaperI-IRT-twophon} Current (top) and conductance (bottom) for
 the model with two vibrational modes 
in the resonant 
transport regime. 
The dashed black line refers to a 
purely electronic calculation. The solid gray line depicts results with 
vibrations in thermal equilibrium, $\mathbf{\Pi}_{\text{el}}=0$, while the solid black 
line presents results including nonequilibrium effects, 
$\mathbf{\Pi}_{\text{el}}\neq0$.}
\end{figure}

We next consider the resonant transport regime, which is found for voltages 
$\Phi\gtrsim1.5\unit{V}$. Fig.\ \ref{PaperI-IRT-twophon} shows the 
current and the conductance in this regime based on three different calculations:
full vibronic calculations with (solid black lines) and 
without (solid gray lines) vibrational nonequilibrium effects
as well as results of a purely electronic calculation (i.e.\ $\lambda_{\alpha}=0$, black dashed lines). 
The purely electronic calculation
exhibits a single peak in the conductance at a bias voltage of $\Phi\approx2\epsilon_{0}$ 
when the molecular level 
enters the conductance window  between the chemical potentials of the left and the right lead
($\mu_{\text{L}}>\epsilon_{0}>\mu_{\text{R}}$). The 
subsequent decrease of the current results from the finite width of the 
conduction band.\cite{Cizek04}

\begin{figure}
\resizebox{0.85\textwidth}{0.56\textwidth}{
\includegraphics{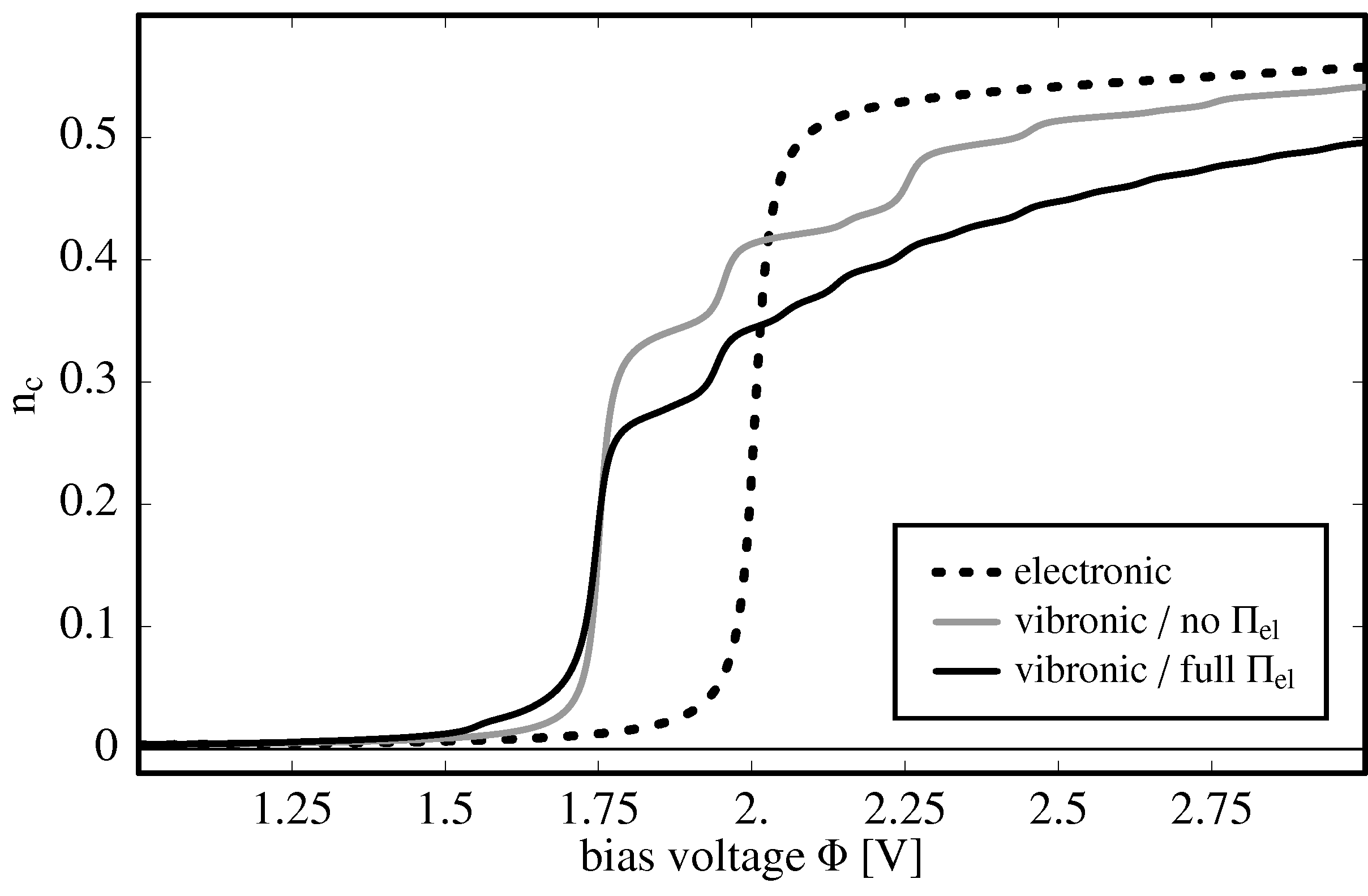}
}\\
\caption{\label{PaperI-nFocc-twophon} Population of the electronic state, $n_{c}=\im{G_{c}^{<}(t=0)}$, for
 the model with two vibrational modes.}
\end{figure}

The coupling to the vibrational degrees of freedom manifests 
itself in steps in the current  and as peaks in the conductance. 
For the present model ($\delta = 0$), these resonance structures can be approximately
associated to transitions between the vibrational states of the neutral molecule and those
of the molecular anion.
Thereby, transitions from vibrational states of the neutral molecule to those of the anion dominate
in the low-voltage regime, where the electronic state on the molecular bridge is essentially unoccupied (cf. Fig.\ \ref{PaperI-nFocc-twophon}).
For higher voltages, the electronic population on the molecular bridge is no longer
negligible and, therefore, also transitions from the vibrational states in the molecular anion to
those in the neutral system contribute to the current. Because, in the present model, 
the vibrational frequencies are the same for both electronic states, features related to these transitions
appear at the same voltages and cannot straightforwardly be distinguished. The population of the electronic state is depicted in Fig.\ \ref{PaperI-nFocc-twophon}. The electronic population increases with bias voltage in a step-like fashion similar to the current-voltage characteristic (cf. Fig.\ \ref{PaperI-IRT-twophon}(a)). In contrast to the current, which decreases for large voltages due to the finite band width, the population increases for larger voltages and exceeds the wide-band limit, which is $0.5$ for a symmetric junction.

If  vibrational 
nonequilibrium effects are neglected ($\mathbf{\Pi}_{\text{el}} = 0$, solid gray line),
the molecule is (due to the low temperature) in its vibrational ground state. 
Correspondingly, the transmitting electrons can only induce transitions starting from
the vibrational ground state.
The lowest peak  in the conductance corresponds predominantly
to the transition from the vibrational ground state in the neutral molecule
to the vibrational ground state of the anion. 
Due to the polaron shift (cf.\ Eq.\ (\ref{polaron_shift})), which accounts
for the difference between the vertical (purely electronic) and adiabatic transition energy,
this peak appears  at a 
lower bias voltage ($\Phi\approx2\overline{\epsilon}_{0}$, with 
$\overline{\epsilon}_{0}=0.874\unit{eV}$) than the peak in the purely electronic calculation.
According to the moderate coupling 
parameters, $\lambda_{1(2)}/\Omega_{1(2)}=0.6$, this conductance peak has the largest intensity. 
Several resonance structures appear at larger voltages. These structures can be approximately
associated with vibrationally excited states in the molecular anion. The relative peak heights follow qualitatively the respective Franck-Condon factors. However, the intensities do not coincide with the relative Franck-Condon factors because both electron and hole transport contribute to the current.

If  vibrational nonequilibrium effects are included, $\mathbf{\Pi}_{\text{el}}\neq0$, 
additional peaks appear in  the conductance. 
These are due to the fact that the nonequilibrium stationary state of the vibrational modes
is no longer the ground state but involves excited states.
Thus, in addition to the transitions considered above, where the electron can only
lose energy to the vibrations, the electrons may
induce transitions from vibrationally higher excited states to lower
vibrational states corresponding to the absorption of vibrational energy by the 
transmitting electrons. As a result, the 
current (solid black lines) increases already before 
the shifted electronic state enters the conductance window  
$[\mu_{\text{L}},\mu_{\text{R}}]$. The four conductance peaks, which are 
seen in the inset of  Fig.\ \ref{PaperI-IRT-twophon}(b),  are caused predominantly
by such absorptive 
processes. 
Some of the structures involve both excitation and deexcitation of vibrational modes.
For example, the peak at $\Phi \approx 1.45\unit{V}$ is associated to emission of energy to mode 
 $(1)$ and absorption of energy from  mode $(2)$. 
Similar processes are found for higher voltages. The position 
of features that exist only due to absorptive processes is highlighted by thin dashed lines. For voltages
$\Phi\gtrsim1.75\unit{V}$, the current including vibrational nonequilibrium effects
is found to be smaller than the current neglecting such effects.
This is in agreement with previous model studies.\cite{Mitra04} 
The reduced current reflects the fact that transitions between the neutral molecule and the anion are suppressed for a vibrationally excited molecular bridge.

\begin{figure}
\resizebox{0.85\textwidth}{0.56\textwidth}{
\includegraphics{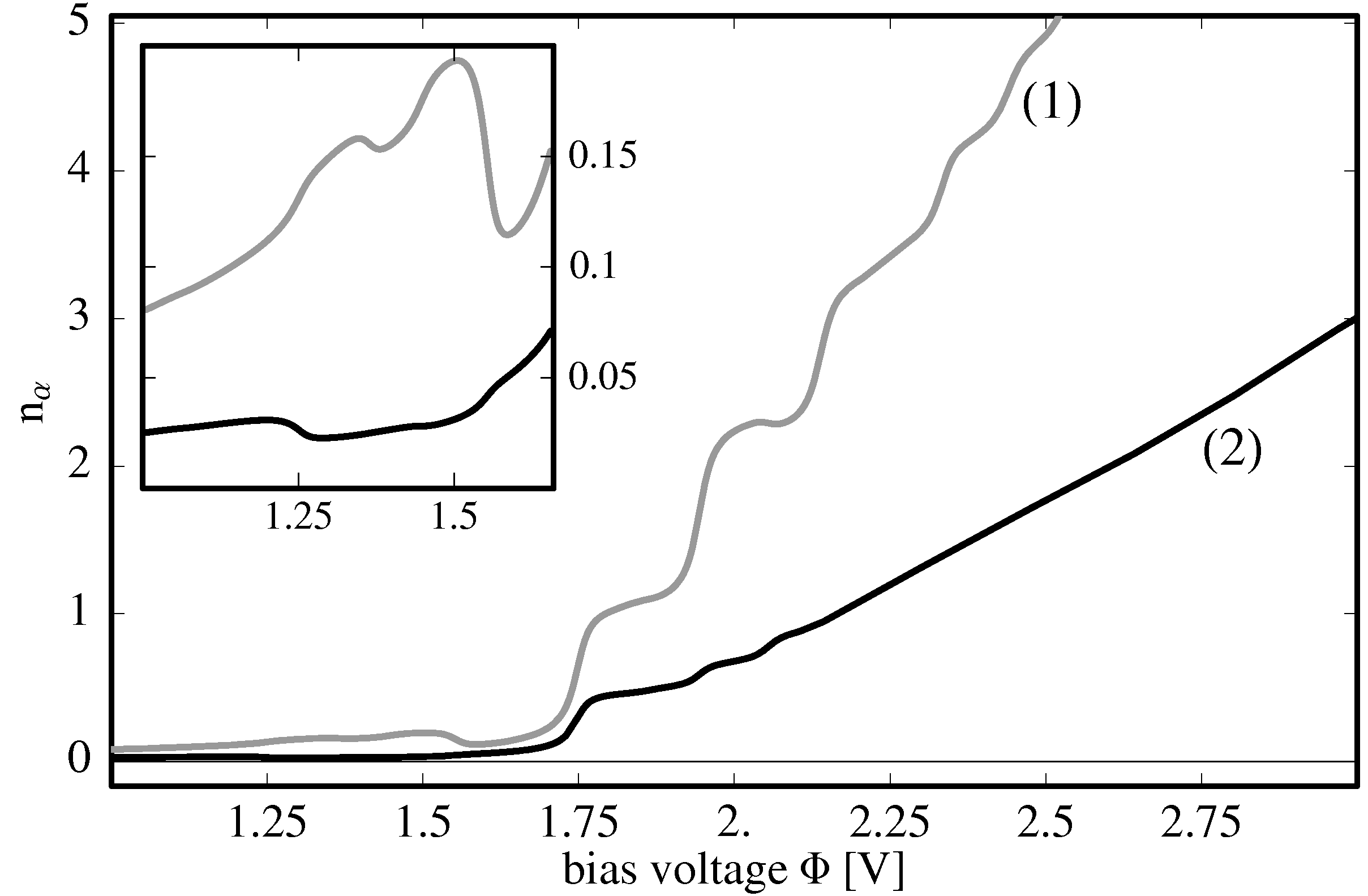}
}\\
\caption{\label{PaperI-nBocc-twophon} Nonequilibrium vibrational excitation for
 the model with two vibrational modes. Solid 
gray lines refer to the mode with $\Omega_{1}=0.10\unit{eV}$, and the solid 
black lines to the one with $\Omega_{2}=0.25\unit{eV}$.}
\end{figure}

The findings discussed above are corroborated by Fig.\ \ref{PaperI-nBocc-twophon}, which 
shows the nonequilibrium vibrational excitation  $n_{1(2)}$ in the stationary state. 
For voltages in the resonant regime ($\Phi > 1.5\unit{V}$), the results exhibit 
pronounced vibrational excitation, corresponding to a  significant deviation of
the nonequilibrium
vibrational distribution  from the equilibrium distribution.
Because excitation of 
mode (1) requires less energy than that of mode (2), and because of the equal coupling 
parameters $\lambda_{1(2)}/\Omega_{1(2)}=0.6$, $n_{1}$ is systematically larger than 
$n_{2}$. The vibrational occupation numbers 
exhibit steplike changes at the same bias voltages as the current.\cite{Mitra04}
In contrast to the current-voltage characteristic, however, 
at some steps the vibrational excitation decreases, corresponding to 
 absorptive processes.

\begin{figure}
\resizebox{0.85\textwidth}{0.56\textwidth}{
\includegraphics{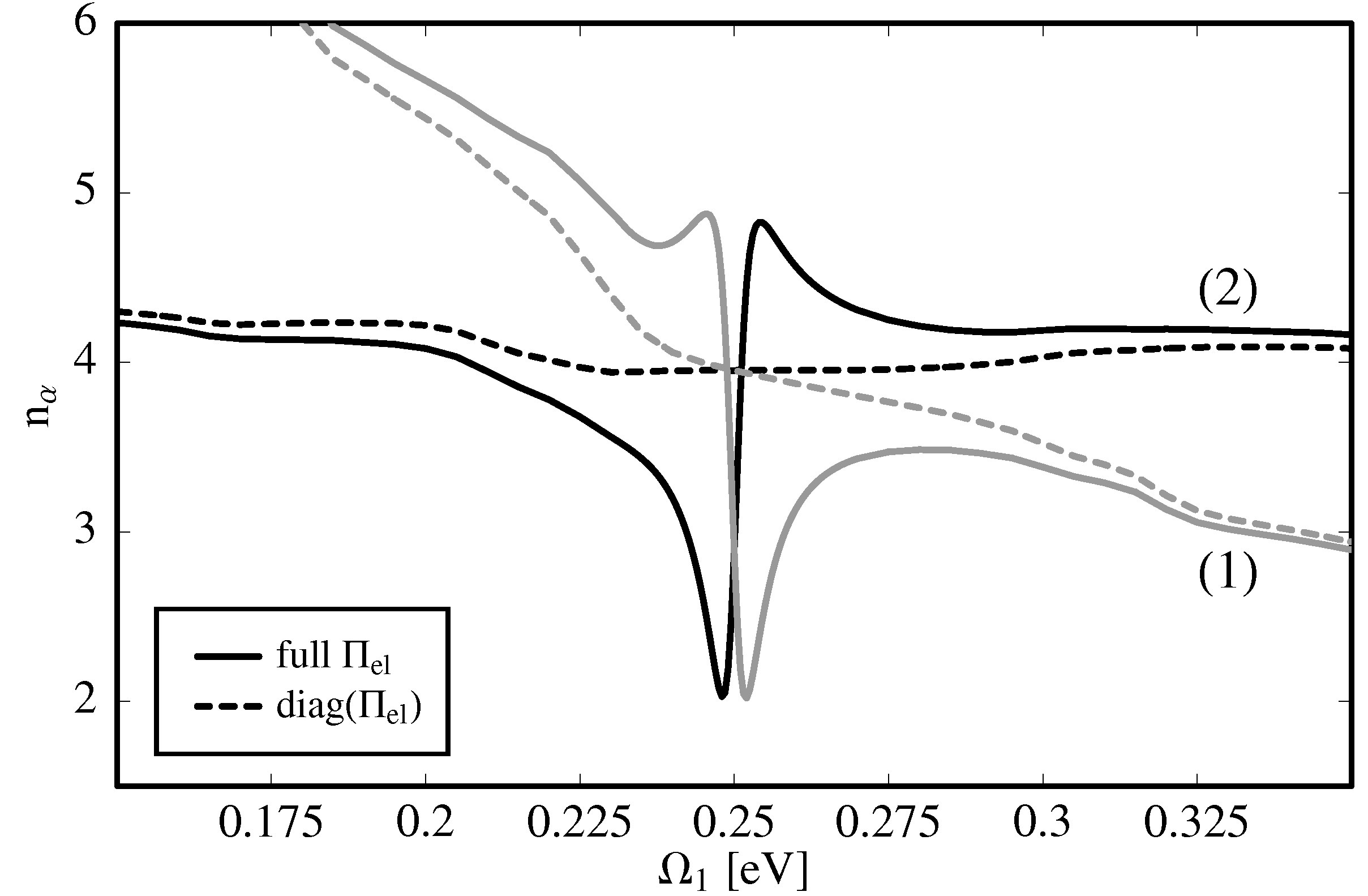}
}\\
\caption{\label{PaperI-nBocc-degeneracy} Vibrational 
excitation numbers $n_{1(2)}$ versus
the vibrational frequency $\Omega_{1}$ for the model with two vibrational modes. The bias voltage is set to 
$3.5\unit{V}$. Gray lines refer to mode (1), black lines 
to mode (2). Dashed lines are calculated neglecting the off-diagonal elements of 
$\mathbf{\Pi}_{\text{el}}$, while the solid lines are obtained using the full 
self energy matrix.}
\end{figure}

Finally, we discuss the coupling of the two vibrational modes mediated
by the electronic degrees of freedom. Technically, this coupling
is described by the off-diagonal elements of the self energy $\mathbf{\Pi}_{\text{el}}$,
which depend both on the electronic molecule-lead coupling and the electronic-vibrational 
coupling.
Because the  coupling of the vibrational modes
is mediated by the electronic degrees of freedom, it
would enter a perturbative description only in second order.
As a result, for the relatively small molecule-lead coupling considered here,
large contributions are only expected for quasidegenerate vibrational modes,
 $\Omega_{1}\approx\Omega_{2}$.
To study this effect, Fig.\ \ref{PaperI-nBocc-degeneracy} shows the nonequilibrium vibrational
excitation  of the two modes for fixed voltage ($\Phi = 3.5\unit{V}$) as a function of the 
frequency of mode (1). Thereby, the electronic-vibrational coupling strength $\lambda_1$ was adjusted
to keep the dimensionless coupling parameter at the constant value $\lambda_{1}/\Omega_{1}=0.6$.
 All other parameters of the model  remain 
unchanged.
In addition to the results including the full self energy $\mathbf{\Pi}_{\text{el}}$
(solid lines), 
results neglecting the off-diagonal elements of $\mathbf{\Pi}_{\text{el}}$ are depicted 
(dashed lines).
The results demonstrate that for vibrational modes with significantly different frequencies, 
as considered above in Figs.\ \ref{PaperI-IRT-twophon}--\ref{PaperI-nBocc-twophon}, 
the electronically mediated mode-mode coupling has little effect and thus the off-diagonal
elements of $\mathbf{\Pi}_{\text{el}}$ could be neglected.
For quasidegenerate modes with similar frequencies $\Omega_{1}\approx\Omega_{2}$, 
on the other hand, neglecting the off-diagonal
elements of $\mathbf{\Pi}_{\text{el}}$ would result in qualitatively incorrect results.
In particular, the 
results including the full self energy matrix
predict a resonant behavior of the vibrational excitation
for quasidegenerate modes, which is  missed if the off-diagonal elements
of the self energy $\mathbf{\Pi}_{\text{el}}$ are neglected.

It is noted, that if the modes are exactly degenerate, $\Omega_{1}$ = $\Omega_{2}$ = $\Omega$, 
and have the same vibronic coupling, $\lambda_{1}$ = $\lambda_{2}$ = $\lambda$, the system 
can be equivalently represented by a single vibrational mode with energy $\Omega$ and coupling 
parameter $\sqrt{2}\lambda$ (or $\sqrt{N}\lambda$ for $N$ degenerate modes).
Calculating the 
excitation number $n_{\text{\tiny single}}$ for this single mode  system, 
we obtain $n_{\text{\tiny single}}\approx n_{1}+n_{2}=2n_{1(2)}$. The
agreement of $n_{\text{\tiny single}}$ with $2n_{1(2)}$ corroborates the approximations employed in the derivation of Eq.\ (\ref{nBocc}).

\subsection{Benzenedibutanethiolate}

As a second application, we consider electron transport through
$p$-benzene-di(butanethiolate) (BDBT) bound to gold electrodes.
This system was chosen because the butyl spacer group acts as an insulator
between the electronic $\pi$-system of benzene and the gold electrodes.
Compared to benzenedithiolate,\cite{Benesch06} the residence time of the electron on
the molecular bridge is thus significantly longer, which results in pronounced vibrational
effects. In recent work, we have developed a first-principles model 
to describe vibrationally-coupled electron transport through BDBT and
studied conductance properties employing inelastic scattering theory.\cite{note_Benesch08b}
Here, we apply the nonequilibrium Green's function approach outlined above to
investigate charge transport through BDBT.

The details of the model are described in Ref.\ \onlinecite{note_Benesch08b}.
Briefly, electron transport through BDBT is dominated by two electronic states localized
at the molecular bridge (denoted A and B in the following). The energies of these two  
states are located at   $\epsilon_{\text{A}}=-1.38\unit{eV}$, 
$\epsilon_{\text{B}}=-1.77\unit{eV}$ with respect to the Fermi energy of the 
junction. As a result, the states are occupied in equilibrium, corresponding to $\delta=1$.
The model includes the four most strongly coupled vibrational modes. The parameters of these modes
are given in Tab.\ \ref{parfourvib}. Vibrational relaxation is described in 
the same manner as for the model system considered above. 

\begin{table}
\begin{tabular}{|ccc|ccc|ccc|ccc|ccc|}
\hline
&frequency&&&vibronic coupling in state A &&& vibronic coupling in state B &&&system-bath coupling& \\
\hline
&$\Omega_{\text{a}}=0.070\unit{eV}$&&&$\lambda_\text{a}^{(\text{A})}=0.021\unit{eV}$&&&$\lambda_\text{a}^{(\text{B})}=0.049\unit{eV}$&&&$
\eta_{\text{a}}=0.001$&\\
&$\Omega_{\text{b}}=0.149\unit{eV}$&&&$\lambda_\text{b}^{(\text{A})}=0.052\unit{eV}$&&&$\lambda_\text{b}^{(\text{B})}=0.037\unit{eV}$&&&$
\eta_{\text{b}}=0.001$&\\
&$\Omega_{\text{c}}=0.153\unit{eV}$&&&$\lambda_\text{c}^{(\text{A})}=0.039\unit{eV}$&&&$\lambda_\text{c}^{(\text{B})}=0.080\unit{eV}$&&&$
\eta_{\text{c}}=0.001$&\\
&$\Omega_{\text{d}}=0.208\unit{eV}$&&&$\lambda_\text{d}^{(\text{A})}=0.120\unit{eV}$&&&$\lambda_\text{d}^{(\text{B})}=0.093\unit{eV}$&&&$
\eta_{\text{d}}=0.001$&\\
\hline
\end{tabular}
\caption{\label{parfourvib} Vibrational parameters for the benzenedibutanethiolate molecular junction.}
\end{table}

Strictly speaking, the nonequilibrium approach outlined above can only be applied
to a single electronic state on the molecular bridge. The extension
of the approach to allow the description of multiple electronic
states will be the subject of future work.
In the system considered here, the electronic states are well separated 
from each other (with respect to the corresponding level broadening, 
$\Gamma_{\text{K}}$), and, furthermore,  at least three vibrational quanta are required to 
bridge the electronic energy gap,  $\overline{\epsilon}_{\text{A}}-\overline{\epsilon}_{\text{B}}\approx0.37\unit{eV}$. 
Therefore, we neglect coherences between the two electronic states and calculate
the overall current as the sum of
the currents obtained separately for the two states. This approximate treatment
is supported by purely electronic as well as inelastic scattering theory
calculations that include both electronic states simultaneously.\cite{note_Benesch08b}

\begin{figure}
\resizebox{0.85\textwidth}{0.56\textwidth}{
\includegraphics{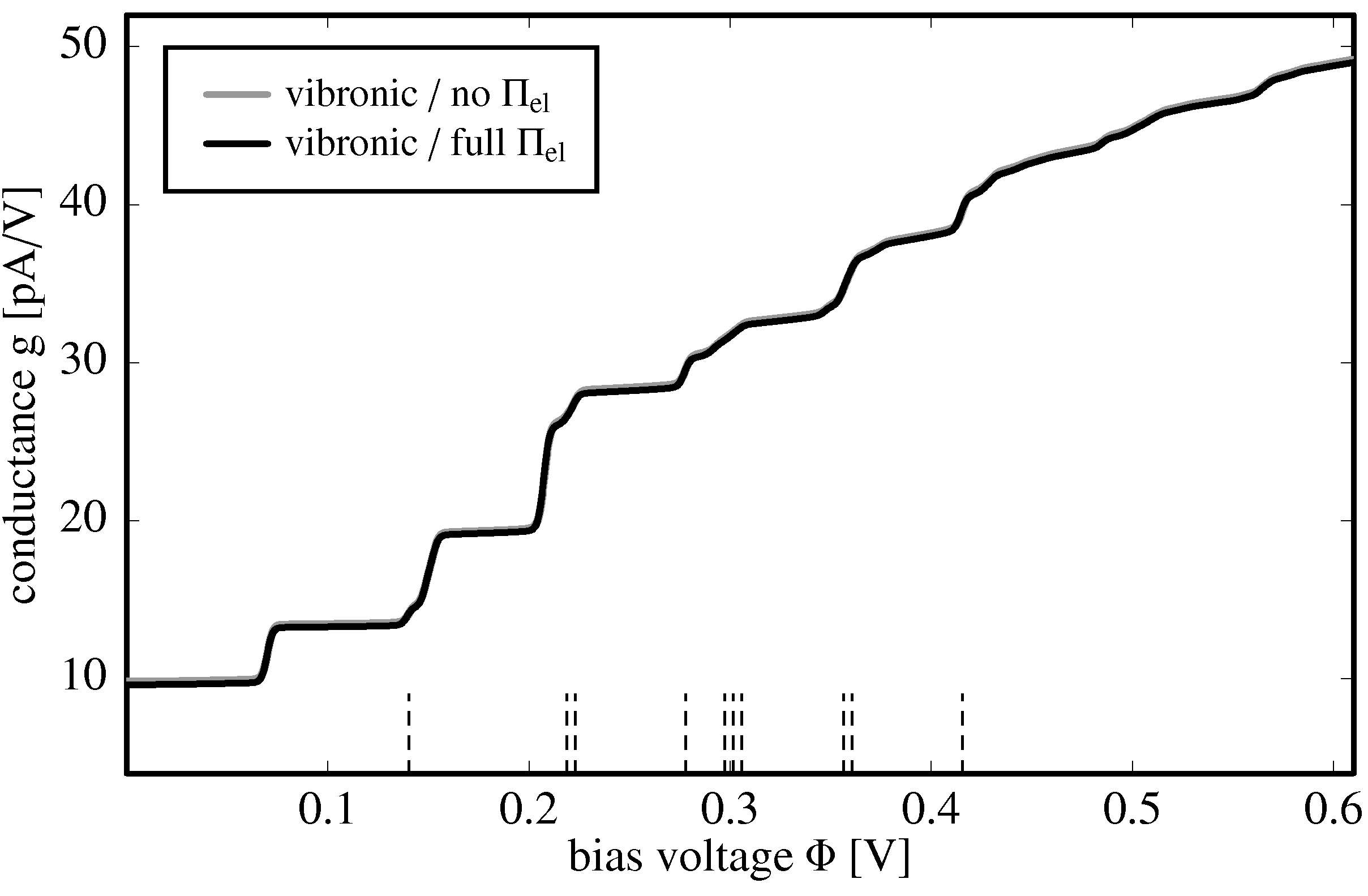}
}\\
\caption{\label{IETS-butyl} Conductance of a benzenedibutanethiolate molecular junction 
in the inelastic 
tunneling regime. The solid gray line refers to a calculation with vibrations 
in thermal equilibrium, $\mathbf{\Pi}_{\text{el}}=0$, while the solid black 
line presents results including nonequilibrium effects, 
$\mathbf{\Pi}_{\text{el}}\neq0$.}
\end{figure}

We start our discussion with the conductance in the inelastic tunneling regime, 
shown in Fig.\ \ref{IETS-butyl}. The results exhibit distinct steps at the onset of the
inelastic channels corresponding to the excitation of a single vibrational quantum, 
$\Phi=\Omega_{\text{a}},\Omega_{\text{b}},..$. 
For larger voltages, structures  related to double excitation processes can be seen. 
These are highlighted with thin dashed lines.
The comparison of the results with and without self energy $\mathbf{\Pi}_{\text{el}}$
shows that the renormalization of the vibrational 
energies, originating from interactions with the electronic degrees of freedom, is negligible.

\begin{figure}
\resizebox{0.85\textwidth}{0.56\textwidth}{
\includegraphics{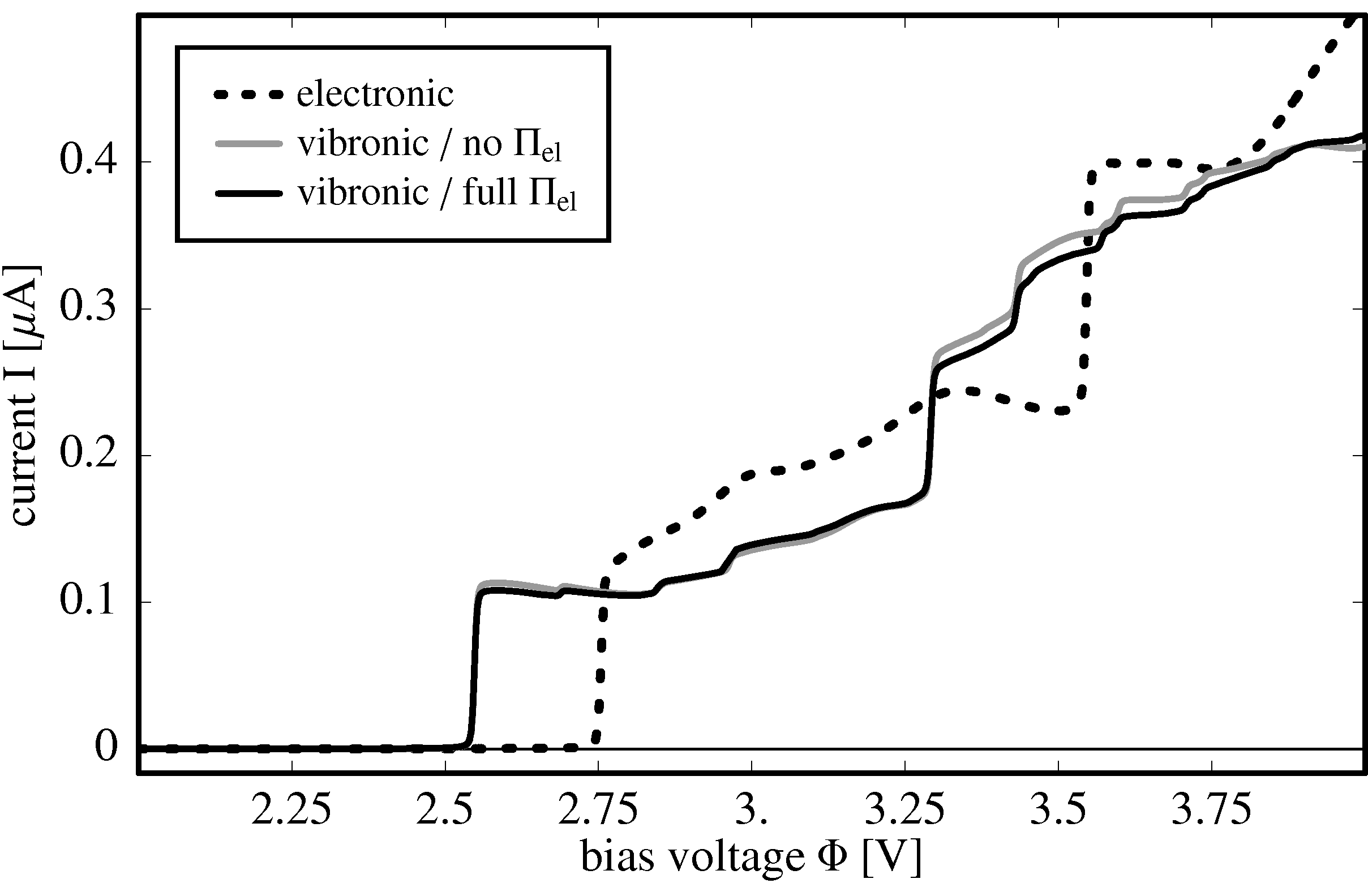}
}\\
\resizebox{0.85\textwidth}{0.56\textwidth}{
\includegraphics{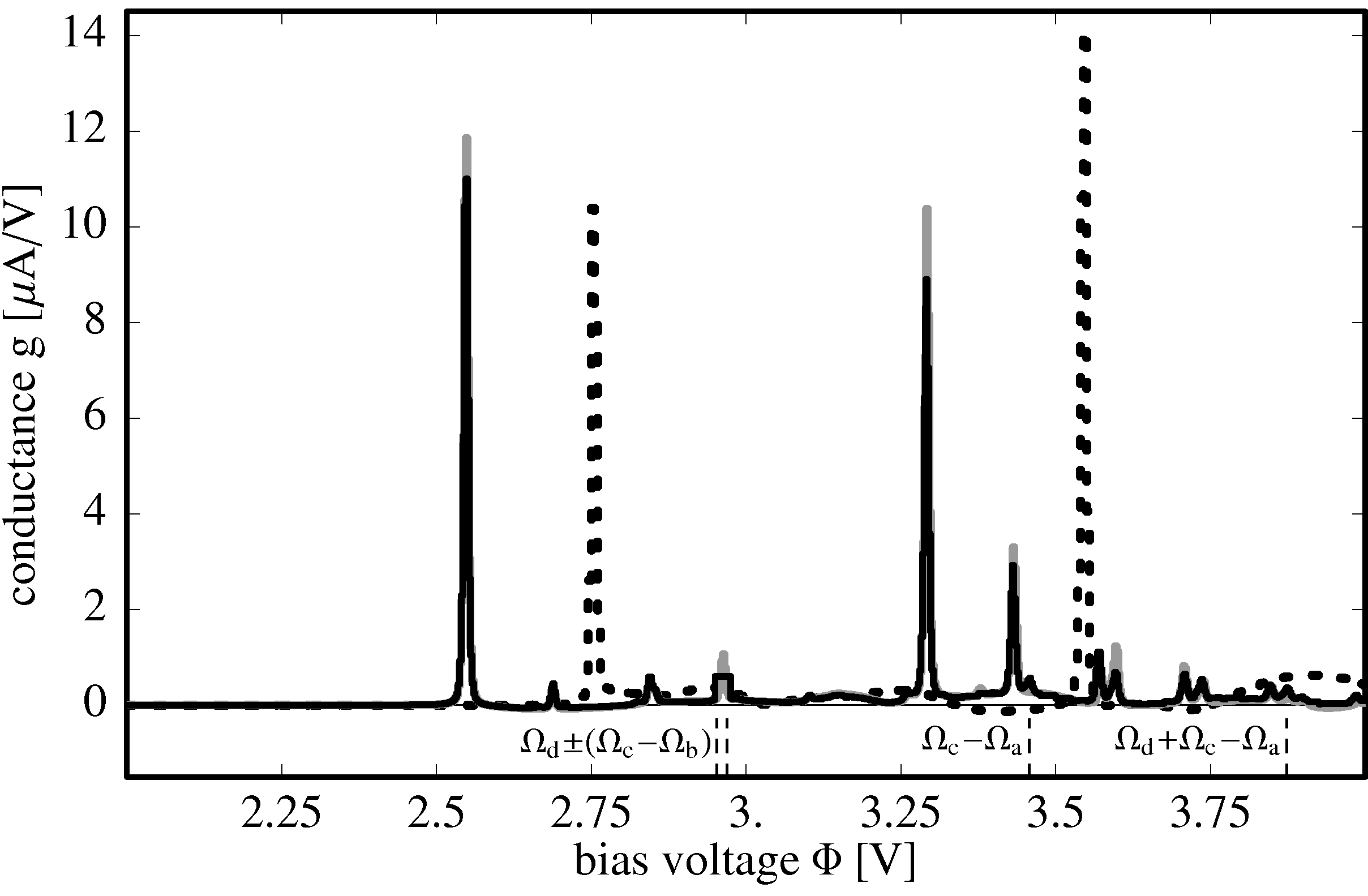}
}\\
\caption{\label{IRT-butyl} Current and conductance of the benzenedibutanethiolate 
molecular junction  in the 
resonant transport regime. The dashed black lines represent results of a purely electronic 
calculation ($\lambda_{\alpha}=0$).
Solid gray lines refer to a calculation with vibrations 
in thermal equilibrium, $\mathbf{\Pi}_{\text{el}}=0$, while solid black 
lines present results including nonequilibrium effects, 
$\mathbf{\Pi}_{\text{el}}\neq0$.}
\end{figure}

Next, we consider charge transport through BDBT in the resonant regime. Fig.\ \ref{IRT-butyl}
depicts results of calculations with and without vibrational nonequilibrium effects.
In addition, results of purely electronic calculations ($\lambda_{\alpha}=0$) are shown.
The purely electronic result (dashed line) shows two pronounced steps at voltages
where the two electronic states enter the conductance window, respectively. All other smaller
structures are due to the energy dependence of the electronic self energy.\cite{note_Benesch08b}

The inclusion of electronic-vibrational coupling results in
a  shift of the two major steps from 
$\Phi=2.76\unit{V}$ to $\Phi=2.54\unit{V}$ and 
from $\Phi=3.54\unit{V}$ to $\Phi=3.29\unit{V}$.
This shift corresponds to the nuclear reorganization energy in the two electronic states.
Furthermore, a number of additional structures appears. For the model considered here,
which is dominated by hole transport through occupied states ($\delta = 1$), 
these  structures can be associated with transitions from vibrational states in
the neutral molecule to those in the molecular cation. The first four peaks
after the onset of the current can be associated with transitions
from the vibrational ground state in the neutral molecule to
singly excited vibrational states in the cation (i.e.\ 'emission' processes) corresponding to
state A. 
The fourth  of these peaks shows a significant broadening. We devote this broadening 
to two mixed emission and absorption processes with an energy transfer of 
$\Delta E=\Omega_{\text{d}}\pm(\Omega_{\text{c}}-\Omega_{\text{b}})$.
For higher voltages, a number of structures can be seen, which are related
 to vibronic transitions
in state B. Thereby, excitation of single 
vibrational quanta is the dominant process and gives rise to the 
fundamental peaks at  voltages $\Phi=3.43\unit{V}$, $3.59\unit{V}$, $3.60\unit{V}$, $3.71\unit{V}$.
Each of 
these peaks is accompanied by another peak that results from excitation of 
an additional quantum of mode (a) at 
$\Phi=3.57\unit{V}$, $3.73\unit{V}$, $3.74\unit{V}$, $3.85\unit{V}$. This result
demonstrates the
strong coupling of mode (a) to state B (cf. Tab.\ \ref{parfourvib}).
Absorptive channels, related to transition from higher excited vibrational  states  
to lower  vibrational states, 
play only a minor role. Three smaller peaks can be assigned
to such processes:  
The first one coincides with the double emission peak at 
$\Phi=3.57\unit{V}$ and corresponds  to emission of a vibrational quantum with energy 
$\Omega_{\text{d}}$ and absorption of a quantum with energy 
$\Omega_{\text{a}}$, i.e.\ 
$\Delta E=\Omega_{\text{d}}-\Omega_{\text{a}}$. 
The 
second peak at $\Phi=3.46\unit{V}$ involves emission of a vibrational quantum with energy 
$\Omega_{\text{c}}$ and absorption of a quantum with energy 
$\Omega_{\text{a}}$, i.e.\ $\Delta E=\Omega_{\text{c}}-\Omega_{\text{a}}$. 
Finally, the small feature at 
$\Phi=3.87\unit{V}$ corresponds to a transition with 
$\Delta E=\Omega_{\text{d}}+\Omega_{\text{c}}-\Omega_{\text{a}}$. 

Electronically mediated mode-mode coupling described by the off-diagonal 
elements of the self energy matrix $\mathbf{\Pi}_{\text{el}}$ has negligible influence on the 
results (data not shown). 
In view of the very similar frequencies of modes (b) and (c), this
finding is at first glance surprising.
 The difference of their frequencies, $\Omega_{\text{c}}-\Omega_{\text{b}}$, is, however,
large compared  to 
the electronic level broadening $\Gamma_{\text{K}}$ thus effectively reducing the 
mode-mode coupling.

\begin{figure}
\resizebox{0.85\textwidth}{0.56\textwidth}{
\includegraphics{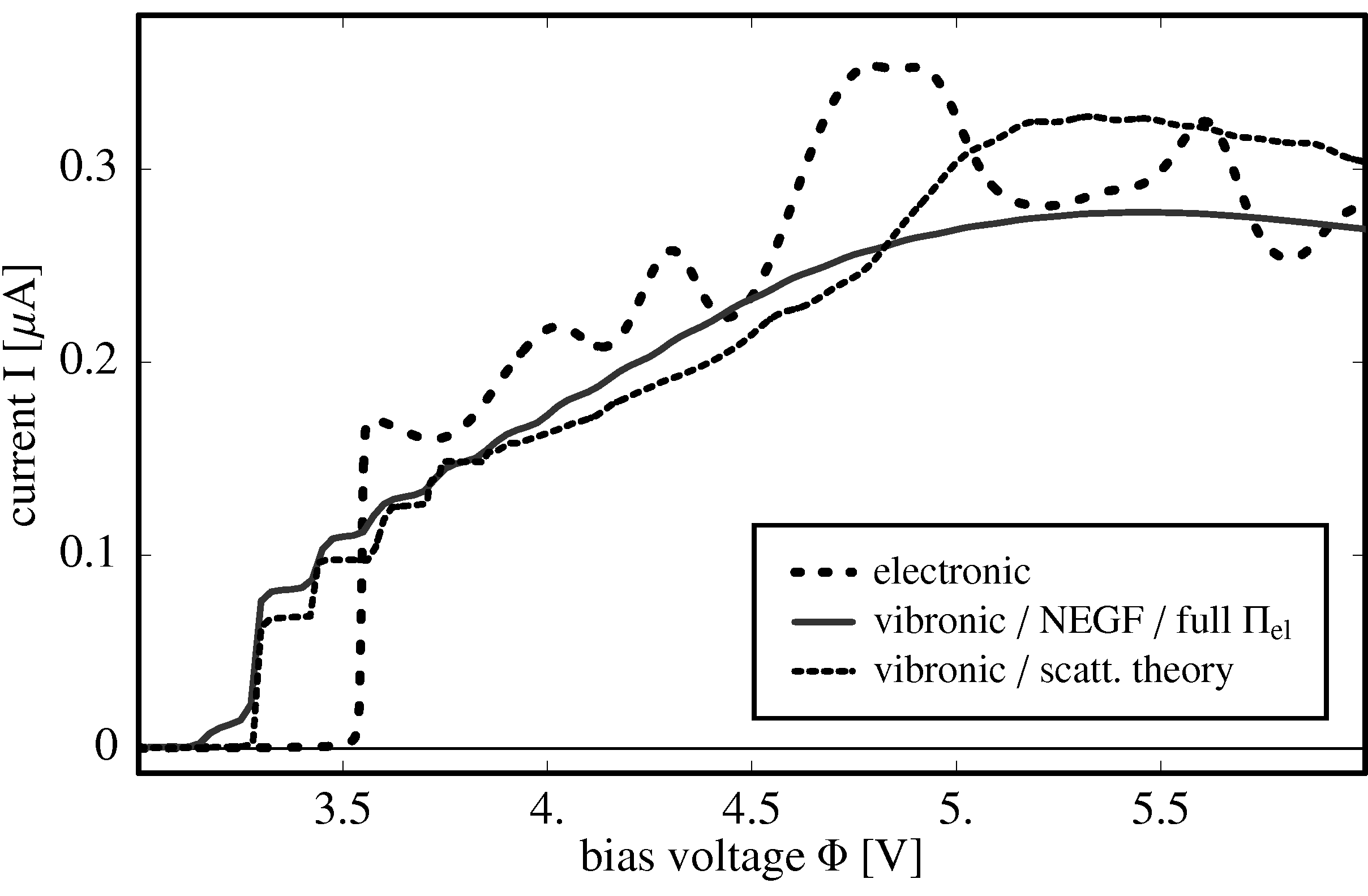}
}\\
\caption{\label{comptoscatttheory} Comparison of different methods for the current 
 through a  benzenedibutanethiolate 
molecular junction. 
Shown are results of vibronic calculations employing the 
nonequilibrium Green's function  approach  
with the full self energy matrix $\mathbf{\Pi}_{\text{el}}$ (solid line)
and inelastic scattering theory (dotted line).\cite{note_Benesch08b}  The dashed 
line depicts results of a purely electronic calculation, where both methods give identical results. 
In contrast to Fig.\ \ref{IRT-butyl}, all calculations include 
only one electronic state (state B).}
\end{figure}

\begin{figure}
\resizebox{0.85\textwidth}{0.56\textwidth}{
\includegraphics{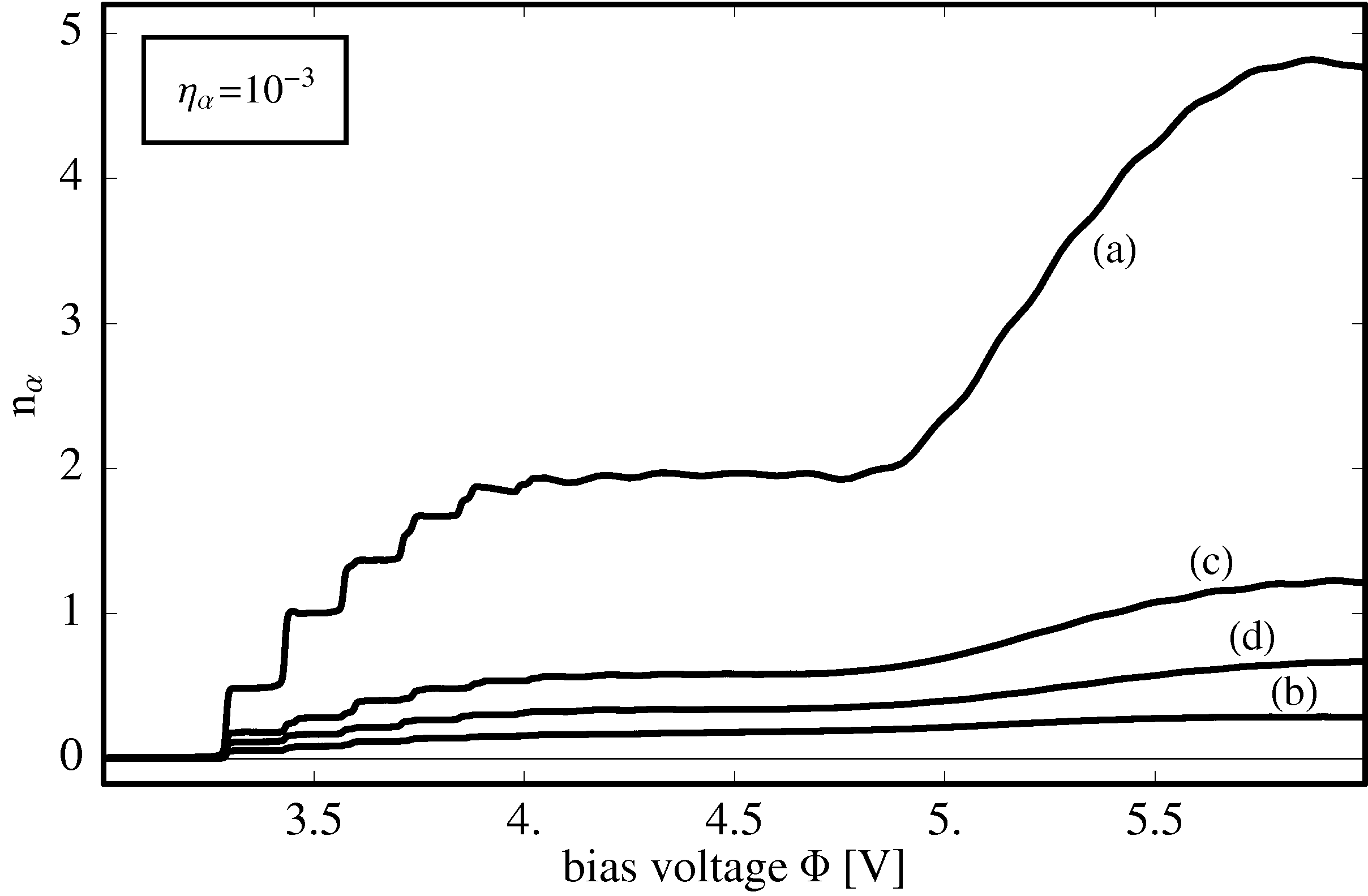}
}\\
\caption{\label{nBocc-butyl-state1234} Nonequilibrium vibrational excitation in 
charge transport through  benzenedibutanethiolate.  Shown is the average number of 
vibrational quanta in the stationary state for modes 
(a)--(d).   The calculation includes only transport through state B   and was obtained for a 
system-bath coupling strength of 
$\eta_{\alpha}=10^{-3}$.}
\end{figure}
\begin{figure}
\resizebox{0.85\textwidth}{0.56\textwidth}{
\includegraphics{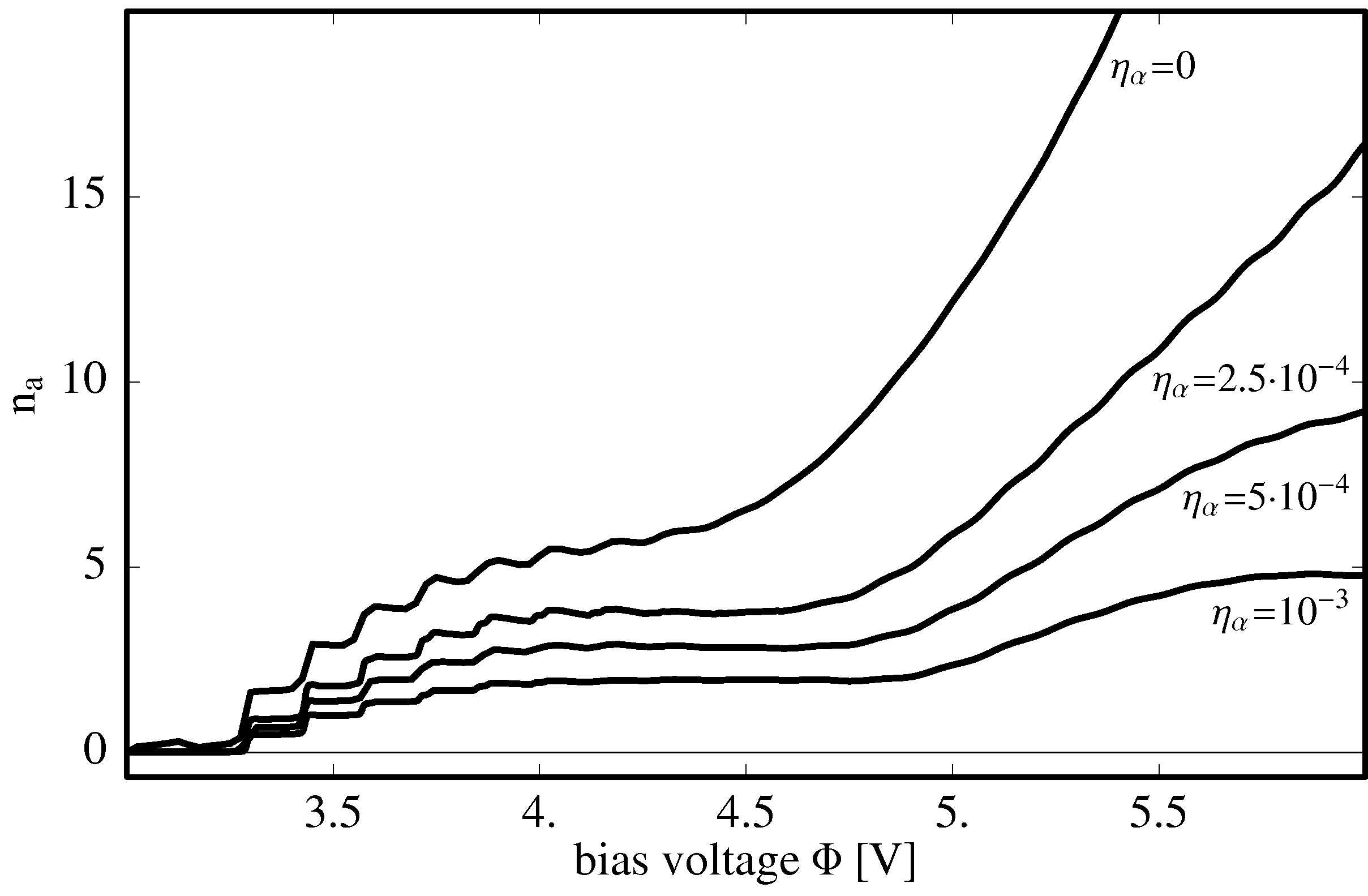}
}\\
\caption{\label{nBocc-butyl-state1}  Nonequilibrium vibrational excitation in 
charge transport through  benzenedibutanethiolate.
Shown is the average number of 
vibrational quanta in the stationary state for mode (a) for 
different system-bath coupling strengths, $\eta_{\alpha}$.
The calculation includes only transport through state B.}
\end{figure}

In a recent study, we have investigated vibrational effects in charge transport through BDBT
employing inelastic scattering theory.\cite{note_Benesch08b} 
Inelastic scattering has the advantage that the transmission process of a single
electron is described numerically exactly.
The calculation of the current, however,  involves the assumption
that the vibrational degrees of freedom of the molecular bridge relax to their equilibrium
distribution between two consecutive electron transmission processes. Furthermore, 
the scattering theory approach cannot account straightforwardly 
for the nonstationary electronic occupation of the molecular states and, therefore, 
fails if the molecular levels, which determine the transport, are located close to the 
Fermi energy.\cite{Cizek04,note_Benesch08b}
Fig.\ \ref{comptoscatttheory} 
shows a comparison of results
obtained for the current through BDBT employing scattering theory and the 
nonequilibrium Green's function approach. In addition, results of a purely electronic
calculation are shown, where both methods give identical results. To allow a direct comparison, 
the results have been obtained 
including only  one electronic state on the molecule (state B)
 and without coupling to the vibrational bath ($\eta_{\alpha}=0$).
Overall, the  results show a rather good agreement between the two methods.
This is due to the fact that the molecule-lead coupling is rather weak and the 
energy of state B is located well below the Fermi energy.
The major differences are an earlier onset of the current and a smaller 
current for larger voltages in the Green's function calculation, as well as different heights 
of the
step structures.
The earlier onset of the current as well as the different step heights are 
caused by absorptive vibrational transitions
related to the nonequilibrium vibrational distribution and 
contributions from electron transport.
Such processes 
are not accounted for in scattering theory, which assumes
the vibrational degrees of freedom to relax to their equilibrium
distribution between two consecutive electron transmission events and (in this case) 
only considers hole transport.
Because the transmission probability is normalized to unity, both 
results approach each other for larger biases ($\Phi\sim3.7\unit{V}$). 
Increasing the bias voltage even further, the results start to deviate again. 
In this regime, current-induced vibrational excitations cannot be neglected and lead to a substantial 
suppression of the current in the Green's function calculation.\cite{Mitra04}

To conclude this section, 
we consider nonequilibrium vibrational excitation induced by the current 
through the BDBT junction. Fig.\ \ref{nBocc-butyl-state1234} shows the average 
vibrational excitation of the four modes. It is seen that in the resonant transport regime
the current results in significant deviations from the equilibrium distribution. 
Overall, the amount of excitation of the four modes follows the value of the 
dimensionless vibronic coupling
$\lambda_{\alpha}/\Omega_{\alpha}$. In particular mode (a),
which has the strongest vibronic coupling, acquires pronounced excitations in the stationary state.

Current-induced excitation of vibrational modes competes with 
relaxation of vibrational energy due to system-bath coupling and thus depends on 
 the timescale of
vibrational relaxation. The influence of the relaxation time is
 illustrated  in 
Fig.\ \ref{nBocc-butyl-state1} for the strongest coupled mode (a). 
The results obtained for  different values of the system-bath 
coupling strength, $\eta_{\alpha}$, show the increase of
vibrational excitation for longer vibrational relaxation times.
In the limit of negligible vibrational relaxation ($\eta_{\alpha} = 0$), mode (a) is excited to rather
high quantum numbers ($n_{\text{a}}> 20$).
The corresponding energy is still smaller than typical
dissociation energies of a C--C or a C--H bond ($E_{\text{diss}}>3\unit{eV}$).
In this regime, however, the harmonic approximation ceases to be valid and 
anharmonicities should be taken into account.

\section{Conclusions}

In this paper we have studied the effect of multimode vibrational dynamics on 
charge transport through single molecule junctions.
To this end,
we have extended a nonequilibrium Green's function  method, developed by 
Galperin et al.,\cite{Galperin06}
to treat multiple vibrational modes in transport calculations. 
This method is based on a polaron transformation of the Hamiltonian and  
employs perturbation theory within a self-consistent scheme to solve the equations of motion
for the nonequilibrium Green's function. In addition to the simulation
of electronic properties such as the electronic current and the conductance, we have also
outlined a scheme to calculate
the average vibrational excitation in the stationary state
using the nonequilibrium Green's function approach.

The methodology has been applied to two examples: a generic model of
a molecular junction with two active vibrational modes as well as charge transport through 
\text{benzenedibutanethiolate} covalently bound to gold electrodes based on a first-principles model.
In both cases, the results
show that the electronic-vibrational coupling may have significant effects on the
current through the molecular junction. 
The coupling to the vibrational degrees of freedom manifests itself in 
pronounced structures in the current-voltage characteristic and the differential conductance.
Moreover, the current-induced excitation of vibrational modes may
result in a significant deviation of the nonequilibrium vibrational distribution from
the equilibrium distribution.
For modes with similar frequency, mode-mode coupling mediated by the electronic
degrees of freedom is of importance. The results show that in this case
the full self energy matrix needs to be taken into account in the calculation.

The study in this paper was based on a model, which describes the
vibrational degrees of freedom in the harmonic approximation.
This approximation is well suited for small amplitude motion and low excitation energies.
To investigate the influence of higher vibrational excitation and the 
possible dissociation of the molecular bridge,
the approach has to be extended to include anharmonic effects. 
This will be the subject of future work.

\section{Acknowledgment}
We thank Martin Cizek and Wolfgang Domcke for helpful discussions.
This work has been supported by the Deutsche Forschungsgemeinschaft, 
a Grant from the German-Israeli Foundation for Scientific Development (G.I.F.), 
and the Fonds der chemischen Industrie. 
Generous allocation of computing time by the Leibniz Rechenzentrum, Munich,
is gratefully acknowledged.

\appendix

\section{Average number of vibrational excitations}
\label{appA}

In this Appendix, we outline the method used to calculate
the average vibrational excitation in the stationary state
\begin{eqnarray}
n_{\alpha} &=& \mean{a^{\dagger}_{\alpha} a_{\alpha}}_{H}.
\end{eqnarray}
In contrast to the population of the electronic state $n_{c}$, the 
excitation number $n_{\alpha}$ of mode $\alpha$ is not an invariant of the 
Lang-Firsov transformation,
\begin{eqnarray}
n_{c} &=& \mean{c^{\dagger}c}_{H} = \mean{ c^{\dagger} X 
X^{\dagger} c }_{\overline{H}} = \mean{ c^{\dagger} c }_{\overline{H}} = \im{G_{c}^{<}(t=0)},  \\
\label{n_alpha}
n_{\alpha} &=& \mean{a^{\dagger}_{\alpha} a_{\alpha}}_{H} = \begin{cases}
\mean{a^{\dagger}_{\alpha}a_{\alpha}}_{\overline{H}} - \frac{\lambda_{\alpha}}{\Omega_{\alpha}} 
\mean{  Q_{\alpha} c^{\dagger} c }_{\overline{H}} + \frac{\lambda_{\alpha}^{2}}{\Omega_{\alpha}^{2}} 
n_{c} ,& \text{$\delta$=0},\\
\mean{a^{\dagger}_{\alpha}a_{\alpha}}_{\overline{H}}
- \frac{\lambda_{\alpha}}{\Omega_{\alpha}} \mean{  Q_{\alpha} 
c^{\dagger} c }_{\overline{H}} + \frac{\lambda_{\alpha}^{2}}{\Omega_{\alpha}^{2}} \left( 1 - 
n_{c}\right) + \frac{\lambda_{\alpha}}{\Omega_{\alpha}} \mean{  Q_{\alpha} }_{\overline{H}},& \text{$\delta$=1.}\\
\end{cases} 
\end{eqnarray}
As a result, we have to calculate four different expectation values to obtain the 
average excitation number $n_{\alpha}$. Thereby, the subscripts $H/\overline{H}$ denote the Hamilton-operator that is used to compute the respective expectation values. In the following we consider only expectation values calculated with $\overline{H}$, for which the corresponding subscript is omitted from now on.

The third term in Eqs.\ (\ref{n_alpha}), 
$\propto\frac{\lambda_{\alpha}^{2}}{\Omega_{\alpha}^{2}}$, which can be interpreted 
as the contribution from polaron formation, can directly be extracted from the 
electronic Green's function $G_{c}$. It only contributes in the resonant transport 
regime, where $n_{c}$ deviates significantly from its equilibrium value $\delta$. 

The fourth  term in Eqs.\ (\ref{n_alpha}) for the case $\delta=1$ contains a single displacement operator $Q_{\alpha}$. 
For its determination we consider the steady state relations
\bml
\begin{eqnarray}
0 &=& \frac{i\partial \mean{\textbf{P}_{\text{a}}}}{\partial t} = - i 
\textbf{W}_{\text{a}} \mean{ \textbf{Q}_{\text{a}} } -2i \textbf{U}  
\mean{ \textbf{Q}_{\text{b}} },\\
0 &=& \frac{i\partial \mean{\textbf{P}_{\text{b}}}}{\partial t} = - i 
\textbf{W}_{\text{b}} \mean{ \textbf{Q}_{\text{b}} } -2i \textbf{U}^{\dagger}  \mean{ 
\textbf{Q}_{\text{a}} },
\end{eqnarray}
\eml
which result in 
\begin{eqnarray}
\left(\textbf{1}-4 
\textbf{W}_{\text{a}}^{-1}\textbf{U}\textbf{W}_{\text{b}}^{-1}\textbf{
U}^{\dagger}\right) \mean{\textbf{Q}_{\text{a}}}&=&0.
\end{eqnarray}
Because the eigenvalues 
of the matrix $4 
\textbf{W}_{\text{a}}^{-1}\textbf{U}\textbf{W}_{\text{b}}^{-1}\textbf{
U}^{\dagger}$ have to be smaller than unity in order to perform the Lang-Firsov 
transformation (cf.\ Sec.\ \ref{secondsection}), the 
expectation value of $Q_{\alpha}$ vanishes, $\mean{Q_{\alpha}}=0$.

To treat the second term in Eq.\ (\ref{n_alpha}), we consider 
similar steady-state relations
\bml
\begin{eqnarray}
0 &=& \frac{i\partial \mean{\textbf{P}_{\text{a}}c^{\dagger}c}}{\partial t} = - 
i \textbf{W}_{\text{a}} \mean{ \textbf{Q}_{\text{a}}c^{\dagger}c } -2i 
\textbf{U}  \mean{ \textbf{Q}_{\text{b}}c^{\dagger}c } \nonumber\\
&&\hspace{20mm} - \sum_{k\in\text{L,R}} V_{k} \mean{ \textbf{P}_{\text{a}}c_{k}^{\dagger}c 
X} + \sum_{k\in\text{L,R}} V_{k}^{*} \mean{  \textbf{P}_{\text{a}}
c^{\dagger}c_{k} X^{\dagger} },\\
0 &=& \frac{i\partial \mean{\textbf{P}_{\text{b}}c^{\dagger}c}}{\partial t} = - 
i \textbf{W}_{\text{b}} \mean{ \textbf{Q}_{\text{b}} c^{\dagger}c } -2i 
\textbf{U}^{\dagger}  \mean{ \textbf{Q}_{\text{a}} c^{\dagger}c } \nonumber\\
&&\hspace{20mm} - \sum_{k\in\text{L,R}} V_{k} \mean{ \textbf{P}_{\text{b}}c_{k}^{\dagger}c 
X} + \sum_{k\in\text{L,R}} V_{k}^{*} \mean{  \textbf{P}_{\text{b}}
c^{\dagger}c_{k} X^{\dagger} }.
\end{eqnarray}
\eml
Here, additional terms linear in the molecule-lead coupling $V_{k}$ appear. 
However, because all expectation values are taken at equal times, these terms cancel each other. 
This finding is closely related to the existence of a steady-state transport regime 
that preserves Kirchhoff's law, $I_{\text{L}}=-I_{\text{R}}$. Without the terms 
linear in $V_{k}$, we can follow the same line of argument as for 
$\mean{Q_{\alpha}}$, and obtain $\mean{Q_{\alpha}c^{\dagger}c}=0$.

Finally, to calculate $\mean{a_{\alpha}^{\dagger}a_{\alpha}}$, we  
exploit the momentum correlation matrix $\textbf{D}$:
\begin{eqnarray}\label{aa}
\mean{a_{\alpha}^{\dagger}a_{\alpha}} &=&
\frac{1}{4}  \left( \mean{Q_{\alpha}Q_{\alpha}} - 
\im{\left(\textbf{D}(t=0)\right)_{\alpha\alpha}}\right) -\frac{1}{2}.
\end{eqnarray}
The expectation value 
$\mean{Q_{\alpha}Q_{\alpha}}$ can be rewritten in terms of 
$\mean{P_{\alpha}P_{\alpha}}$ and $\mean{Q_{\beta}Q_{\beta}}$. For that 
purpose, we use the following steady-state relations
\bml
\begin{eqnarray}
0 &=& \frac{i\partial\mean{Q_{\alpha}P_{\alpha}}}{\partial t}  = i 
\Omega_{\alpha} \mean{P_{\alpha}P_{\alpha}} - i \Omega_{\alpha} 
\mean{Q_{\alpha}Q_{\alpha}} - 2 i \sum_{\beta} U_{\alpha\beta} 
\mean{Q_{\alpha}Q_{\beta}},\\
0 &=& \frac{i\partial\mean{Q_{\alpha}P_{\beta}}}{\partial t}  = i 
\Omega_{\alpha} \mean{P_{\alpha}P_{\beta}} - i \omega_{\beta} 
\mean{Q_{\alpha}Q_{\beta}} - 2 i \sum_{\alpha'} U_{\alpha'\beta} 
\mean{Q_{\alpha}Q_{\alpha'}},\\
0 &=& \frac{i\partial\mean{P_{\alpha}Q_{\beta}}}{\partial t}  = i 
\omega_{\beta} \mean{P_{\alpha}P_{\beta}} - i \Omega_{\alpha} 
\mean{Q_{\alpha}Q_{\beta}} - 2 i \sum_{\beta'} U_{\alpha\beta'} 
\mean{Q_{\beta'}Q_{\beta}},
\end{eqnarray}
\eml
where, based on the same argument as before, we have  disregarded terms linear in 
$V_{k}$. 
Using these relations, we obtain to second order in the system-bath coupling  
($O(U_{\alpha\beta}^{2})$)
\begin{eqnarray}\label{qq}
\mean{Q_{\alpha}Q_{\alpha}} &=& \mean{P_{\alpha}P_{\alpha}} + 
\sum_{\beta,\alpha'} \frac{4 \omega_{\beta} U_{\alpha\beta} 
U_{\alpha'\beta}}{\Omega_{\alpha'}(\omega_{\beta}^{2}-\Omega_{\alpha}^{2})} 
\mean{P_{\alpha}P_{\alpha'}} - \sum_{\beta} \frac{4 
U_{\alpha\beta}^{2}}{\omega_{\beta}^{2}-\Omega_{\alpha}^{2}} 
\mean{Q_{\beta}Q_{\beta}},\\
&\approx& \mean{P_{\alpha}P_{\alpha}} \left( 1 + \sum_{\beta} \frac{4 
\omega_{\beta} U_{\alpha\beta}^{2} 
}{\Omega_{\alpha}(\omega_{\beta}^{2}-\Omega_{\alpha}^{2})} \right) - 
\sum_{\beta} \frac{4 
U_{\alpha\beta}^{2}}{\omega_{\beta}^{2}-\Omega_{\alpha}^{2}} 
\mean{Q_{\beta}Q_{\beta}}
 .\nonumber
\end{eqnarray}
Thereby, we disregard off-diagonal contributions $\mean{P_{\alpha}P_{\alpha'}}$,
which turn out to be negligible. This step is corroborated by considering the special case of 
 degenerate modes (cf.\ the discussion at the end of Sec.\ 
\ref{twovibrons}). 

Inserting Eq.\ (\ref{qq}) into Eq.\ (\ref{aa}), we obtain
the final expression for the excitation number
\begin{eqnarray}
n_{\alpha} &=& - \left( A_{\alpha} + \frac{1}{2} \right)  
\im{\left(\textbf{D}(t=0)\right)_{\alpha\alpha}} - 
\left(B_{\alpha}+\frac{1}{2}\right) + 
\frac{\lambda_{\alpha}^{2}}{\Omega_{\alpha}^{2}} \begin{cases}
n_{c}, & \delta=0, \\
1- n_{c}, & \delta=1, \\
\end{cases}
\end{eqnarray}
with
\begin{eqnarray} 
A_{\alpha}=\sum_{\beta} 
\frac{U_{\alpha\beta}^{2}\omega_{\beta}}{\Omega_{\alpha}(\omega_{\beta}^{2}
-\Omega_{\alpha}^{2})},\qquad B_{\alpha}=\sum_{\beta} 
\frac{U_{\alpha\beta}^{2}}{(\omega_{\beta}^{2}-\Omega_{\alpha}^{2})} 
\mean{Q_{\beta}Q_{\beta}}.
\end{eqnarray}
The bath modes remain in thermal equilibrium, for which $\mean{Q_{\beta}Q_{\beta}}=1+2N_{\text{B}}(\omega_{\beta})$ with $N_{\text{B}}(\omega)$ the Bose distribution function.

\newpage


\begin{thebibliography}{50}
\expandafter\ifx\csname natexlab\endcsname\relax\def\natexlab#1{#1}\fi
\expandafter\ifx\csname bibnamefont\endcsname\relax
  \def\bibnamefont#1{#1}\fi
\expandafter\ifx\csname bibfnamefont\endcsname\relax
  \def\bibfnamefont#1{#1}\fi
\expandafter\ifx\csname citenamefont\endcsname\relax
  \def\citenamefont#1{#1}\fi
\expandafter\ifx\csname url\endcsname\relax
  \def\url#1{\texttt{#1}}\fi
\expandafter\ifx\csname urlprefix\endcsname\relax\def\urlprefix{URL }\fi
\providecommand{\bibinfo}[2]{#2}
\providecommand{\eprint}[2][]{\url{#2}}

\bibitem[{\citenamefont{Reed et~al.}(1997)\citenamefont{Reed, Zhou, Muller,
  Burgin, and Tour}}]{Reed97}
\bibinfo{author}{\bibfnamefont{M.~A.} \bibnamefont{Reed}},
  \bibinfo{author}{\bibfnamefont{C.}~\bibnamefont{Zhou}},
  \bibinfo{author}{\bibfnamefont{C.~J.} \bibnamefont{Muller}},
  \bibinfo{author}{\bibfnamefont{T.~P.} \bibnamefont{Burgin}},
  \bibnamefont{and} \bibinfo{author}{\bibfnamefont{J.~M.} \bibnamefont{Tour}},
  \bibinfo{journal}{Science} \textbf{\bibinfo{volume}{278}},
  \bibinfo{pages}{252} (\bibinfo{year}{1997}).

\bibitem[{\citenamefont{Reichert et~al.}(2002)\citenamefont{Reichert, Ochs,
  Beckmann, Weber, Mayor, and Lohneysen}}]{Reichert02}
\bibinfo{author}{\bibfnamefont{J.}~\bibnamefont{Reichert}},
  \bibinfo{author}{\bibfnamefont{R.}~\bibnamefont{Ochs}},
  \bibinfo{author}{\bibfnamefont{D.}~\bibnamefont{Beckmann}},
  \bibinfo{author}{\bibfnamefont{H.~B.} \bibnamefont{Weber}},
  \bibinfo{author}{\bibfnamefont{M.}~\bibnamefont{Mayor}}, \bibnamefont{and}
  \bibinfo{author}{\bibfnamefont{H.~v.}~\bibnamefont{Lohneysen}},
  \bibinfo{journal}{Phys. Rev. Lett.} \textbf{\bibinfo{volume}{88}},
  \bibinfo{pages}{176804} (\bibinfo{year}{2002}).

\bibitem[{\citenamefont{Chen et~al.}(2007)\citenamefont{Chen, Hihath, Huang,
  Li, and Tao}}]{Chen07}
\bibinfo{author}{\bibfnamefont{F.}~\bibnamefont{Chen}},
  \bibinfo{author}{\bibfnamefont{J.}~\bibnamefont{Hihath}},
  \bibinfo{author}{\bibfnamefont{Z.}~\bibnamefont{Huang}},
  \bibinfo{author}{\bibfnamefont{X.}~\bibnamefont{Li}}, \bibnamefont{and}
  \bibinfo{author}{\bibfnamefont{N.~J.} \bibnamefont{Tao}},
  \bibinfo{journal}{Annu. Rev. Phys. Chem.} \textbf{\bibinfo{volume}{58}},
  \bibinfo{pages}{535} (\bibinfo{year}{2007}).

\bibitem[{\citenamefont{Selzer and Allara}(2006)}]{Selzer06}
\bibinfo{author}{\bibfnamefont{Y.}~\bibnamefont{Selzer}} \bibnamefont{and}
  \bibinfo{author}{\bibfnamefont{D.~L.} \bibnamefont{Allara}},
  \bibinfo{journal}{Annu. Rev. Phys. Chem.} \textbf{\bibinfo{volume}{57}},
  \bibinfo{pages}{593} (\bibinfo{year}{2006}).

\bibitem[{\citenamefont{{H\"anggi} et~al.}(2002)\citenamefont{{H\"anggi},
  Ratner, and Yaliraki}}]{Haenggi02}
\bibinfo{author}{\bibfnamefont{P.}~\bibnamefont{{H\"anggi}}},
  \bibinfo{author}{\bibfnamefont{M.~A.} \bibnamefont{Ratner}},
  \bibnamefont{and} \bibinfo{author}{\bibfnamefont{S.}~\bibnamefont{Yaliraki}},
  \bibinfo{journal}{Chem.\ Phys.\ , special issue on: {\em "Processes in
  Molecular Wires"}} \textbf{\bibinfo{volume}{281}}, \bibinfo{pages}{111}
  (\bibinfo{year}{2002}).

\bibitem[{\citenamefont{Nitzan and Ratner}(2003)}]{Nitzan03}
\bibinfo{author}{\bibfnamefont{A.}~\bibnamefont{Nitzan}} \bibnamefont{and}
  \bibinfo{author}{\bibfnamefont{M.~A.} \bibnamefont{Ratner}},
  \bibinfo{journal}{Science} \textbf{\bibinfo{volume}{300}},
  \bibinfo{pages}{1384} (\bibinfo{year}{2003}).

\bibitem[{\citenamefont{Cuniberti et~al.}(2005)\citenamefont{Cuniberti, Fagas,
  and Richter}}]{Cuniberti05}
\bibinfo{author}{\bibfnamefont{G.}~\bibnamefont{Cuniberti}},
  \bibinfo{author}{\bibfnamefont{G.}~\bibnamefont{Fagas}}, \bibnamefont{and}
  \bibinfo{author}{\bibfnamefont{K.}~\bibnamefont{Richter}},
  \emph{\bibinfo{title}{Introducing Molecular Electronics}}
  (\bibinfo{publisher}{Springer}, \bibinfo{address}{Heidelberg},
  \bibinfo{year}{2005}).

\bibitem[{\citenamefont{Galperin et~al.}(2007)\citenamefont{Galperin, Ratner,
  and Nitzan}}]{Galperin07}
\bibinfo{author}{\bibfnamefont{M.}~\bibnamefont{Galperin}},
  \bibinfo{author}{\bibfnamefont{M.~A.} \bibnamefont{Ratner}},
  \bibnamefont{and} \bibinfo{author}{\bibfnamefont{A.}~\bibnamefont{Nitzan}},
  \bibinfo{journal}{J. Phys.: Condens. Matter} \textbf{\bibinfo{volume}{19}},
  \bibinfo{pages}{103201} (\bibinfo{year}{2007}).

\bibitem[{\citenamefont{Park et~al.}(2000)\citenamefont{Park, Park, Lim,
  Anderson, Alivisatos, and McEuen}}]{Park00}
\bibinfo{author}{\bibfnamefont{H.}~\bibnamefont{Park}},
  \bibinfo{author}{\bibfnamefont{J.}~\bibnamefont{Park}},
  \bibinfo{author}{\bibfnamefont{A.~K.~L.} \bibnamefont{Lim}},
  \bibinfo{author}{\bibfnamefont{E.~H.} \bibnamefont{Anderson}},
  \bibinfo{author}{\bibfnamefont{A.~P.} \bibnamefont{Alivisatos}},
  \bibnamefont{and} \bibinfo{author}{\bibfnamefont{P.~L.}
  \bibnamefont{McEuen}}, \bibinfo{journal}{Nature (London)}
  \textbf{\bibinfo{volume}{407}}, \bibinfo{pages}{57} (\bibinfo{year}{2000}).

\bibitem[{\citenamefont{Pasupathy et~al.}(2005)\citenamefont{Pasupathy, Park,
  Chang, Soldatov, Lebedkin, Bialczak, Grose, Donev, Sethna, Ralph
  et~al.}}]{Pasupathy05}
\bibinfo{author}{\bibfnamefont{A.~N.} \bibnamefont{Pasupathy}},
  \bibinfo{author}{\bibfnamefont{J.}~\bibnamefont{Park}},
  \bibinfo{author}{\bibfnamefont{C.}~\bibnamefont{Chang}},
  \bibinfo{author}{\bibfnamefont{A.~V.} \bibnamefont{Soldatov}},
  \bibinfo{author}{\bibfnamefont{S.}~\bibnamefont{Lebedkin}},
  \bibinfo{author}{\bibfnamefont{R.~C.} \bibnamefont{Bialczak}},
  \bibinfo{author}{\bibfnamefont{J.~E.} \bibnamefont{Grose}},
  \bibinfo{author}{\bibfnamefont{L.~A.~K.} \bibnamefont{Donev}},
  \bibinfo{author}{\bibfnamefont{J.~P.} \bibnamefont{Sethna}},
  \bibinfo{author}{\bibfnamefont{D.~C.} \bibnamefont{Ralph}},
  \bibnamefont{et~al.}, \bibinfo{journal}{Nano Lett.}
  \textbf{\bibinfo{volume}{5}}, \bibinfo{pages}{203} (\bibinfo{year}{2005}).

\bibitem[{\citenamefont{Gaudioso and Ho}(2001)}]{Gaudioso01}
\bibinfo{author}{\bibfnamefont{J.}~\bibnamefont{Gaudioso}} \bibnamefont{and}
  \bibinfo{author}{\bibfnamefont{W.}~\bibnamefont{Ho}}, \bibinfo{journal}{J.
  Am. Chem. Soc.} \textbf{\bibinfo{volume}{123}}, \bibinfo{pages}{10095}
  (\bibinfo{year}{2001}).

\bibitem[{\citenamefont{Kushmerick et~al.}(2004)\citenamefont{Kushmerick,
  Lazorcik, Patterson, Shashidhar, Seferos, and Bazan}}]{Kushmerick04}
\bibinfo{author}{\bibfnamefont{J.~G.} \bibnamefont{Kushmerick}},
  \bibinfo{author}{\bibfnamefont{J.}~\bibnamefont{Lazorcik}},
  \bibinfo{author}{\bibfnamefont{C.~H.} \bibnamefont{Patterson}},
  \bibinfo{author}{\bibfnamefont{R.}~\bibnamefont{Shashidhar}},
  \bibinfo{author}{\bibfnamefont{D.~S.} \bibnamefont{Seferos}},
  \bibnamefont{and} \bibinfo{author}{\bibfnamefont{G.~C.} \bibnamefont{Bazan}},
  \bibinfo{journal}{Nano Lett.} \textbf{\bibinfo{volume}{4}},
  \bibinfo{pages}{639} (\bibinfo{year}{2004}).

\bibitem[{\citenamefont{Qiu et~al.}(2004)\citenamefont{Qiu, Nazin, and
  Ho}}]{Qiu04}
\bibinfo{author}{\bibfnamefont{X.~H.} \bibnamefont{Qiu}},
  \bibinfo{author}{\bibfnamefont{G.~V.} \bibnamefont{Nazin}}, \bibnamefont{and}
  \bibinfo{author}{\bibfnamefont{W.}~\bibnamefont{Ho}}, \bibinfo{journal}{Phys.
  Rev. Lett.} \textbf{\bibinfo{volume}{92}}, \bibinfo{pages}{206102}
  (\bibinfo{year}{2004}).

\bibitem[{\citenamefont{Djukic et~al.}(2005)\citenamefont{Djukic, Thygesen,
  Untiedt, Smit, Jacobsen, and van Ruitenbeek}}]{Djukic05}
\bibinfo{author}{\bibfnamefont{D.}~\bibnamefont{Djukic}},
  \bibinfo{author}{\bibfnamefont{K.~S.} \bibnamefont{Thygesen}},
  \bibinfo{author}{\bibfnamefont{C.}~\bibnamefont{Untiedt}},
  \bibinfo{author}{\bibfnamefont{R.~H.~M.} \bibnamefont{Smit}},
  \bibinfo{author}{\bibfnamefont{K.~W.} \bibnamefont{Jacobsen}},
  \bibnamefont{and} \bibinfo{author}{\bibfnamefont{J.~M.} \bibnamefont{van
  Ruitenbeek}}, \bibinfo{journal}{Phys. Rev. B} \textbf{\bibinfo{volume}{71}},
  \bibinfo{pages}{161402(R)} (\bibinfo{year}{2005}).

\bibitem[{\citenamefont{Sapmaz et~al.}(2006)\citenamefont{Sapmaz,
  Jarillo-Herrero, Blanter, Dekker, and {van der Zant}}}]{Sapmaz06}
\bibinfo{author}{\bibfnamefont{S.}~\bibnamefont{Sapmaz}},
  \bibinfo{author}{\bibfnamefont{P.}~\bibnamefont{Jarillo-Herrero}},
  \bibinfo{author}{\bibfnamefont{Y.~M.} \bibnamefont{Blanter}},
  \bibinfo{author}{\bibfnamefont{C.}~\bibnamefont{Dekker}}, \bibnamefont{and}
  \bibinfo{author}{\bibfnamefont{H.~S.~J.} \bibnamefont{{van der Zant}}},
  \bibinfo{journal}{Phys. Rev. Lett.} \textbf{\bibinfo{volume}{96}},
  \bibinfo{pages}{026801} (\bibinfo{year}{2006}).

\bibitem[{\citenamefont{Thijssen et~al.}(2006)\citenamefont{Thijssen, Djukic,
  Otte, Bremmer, and van Ruitenbeek}}]{Thijssen06}
\bibinfo{author}{\bibfnamefont{W.~H.~A.} \bibnamefont{Thijssen}},
  \bibinfo{author}{\bibfnamefont{D.}~\bibnamefont{Djukic}},
  \bibinfo{author}{\bibfnamefont{A.~F.} \bibnamefont{Otte}},
  \bibinfo{author}{\bibfnamefont{R.~H.} \bibnamefont{Bremmer}},
  \bibnamefont{and} \bibinfo{author}{\bibfnamefont{J.~M.} \bibnamefont{van
  Ruitenbeek}}, \bibinfo{journal}{Phys. Rev. Lett.}
  \textbf{\bibinfo{volume}{97}}, \bibinfo{pages}{226806}
  (\bibinfo{year}{2006}).

\bibitem[{\citenamefont{{B\"ohler} et~al.}(2007)\citenamefont{{B\"ohler},
  Edtbauer, and Scheer}}]{Boehler07}
\bibinfo{author}{\bibfnamefont{T.}~\bibnamefont{{B\"ohler}}},
  \bibinfo{author}{\bibfnamefont{A.}~\bibnamefont{Edtbauer}}, \bibnamefont{and}
  \bibinfo{author}{\bibfnamefont{E.}~\bibnamefont{Scheer}},
  \bibinfo{journal}{Phys. Rev. B} \textbf{\bibinfo{volume}{76}},
  \bibinfo{pages}{125432} (\bibinfo{year}{2007}).

\bibitem[{\citenamefont{Parks et~al.}(2007)\citenamefont{Parks, Champagne,
  Hutchison, Flores-Torres, Abruna, and Ralph}}]{Parks07}
\bibinfo{author}{\bibfnamefont{J.~J.} \bibnamefont{Parks}},
  \bibinfo{author}{\bibfnamefont{A.~R.} \bibnamefont{Champagne}},
  \bibinfo{author}{\bibfnamefont{G.~R.} \bibnamefont{Hutchison}},
  \bibinfo{author}{\bibfnamefont{S.}~\bibnamefont{Flores-Torres}},
  \bibinfo{author}{\bibfnamefont{H.~D.} \bibnamefont{Abruna}},
  \bibnamefont{and} \bibinfo{author}{\bibfnamefont{D.~C.} \bibnamefont{Ralph}},
  \bibinfo{journal}{Phys. Rev. Lett.} \textbf{\bibinfo{volume}{99}},
  \bibinfo{pages}{026601} (\bibinfo{year}{2007}).

\bibitem[{\citenamefont{Ogawa et~al.}(2007)\citenamefont{Ogawa, Mikaelian, and
  Ho}}]{Ogawa07}
\bibinfo{author}{\bibfnamefont{N.}~\bibnamefont{Ogawa}},
  \bibinfo{author}{\bibfnamefont{G.}~\bibnamefont{Mikaelian}},
  \bibnamefont{and} \bibinfo{author}{\bibfnamefont{W.}~\bibnamefont{Ho}},
  \bibinfo{journal}{Phys. Rev. Lett.} \textbf{\bibinfo{volume}{98}},
  \bibinfo{pages}{166103} (\bibinfo{year}{2007}).

\bibitem[{\citenamefont{Gaudioso et~al.}(2000)\citenamefont{Gaudioso, Lauhon,
  and Ho}}]{Gaudioso00}
\bibinfo{author}{\bibfnamefont{J.}~\bibnamefont{Gaudioso}},
  \bibinfo{author}{\bibfnamefont{L.~J.}~\bibnamefont{Lauhon}}, \bibnamefont{and}
  \bibinfo{author}{\bibfnamefont{W.}~\bibnamefont{Ho}}, \bibinfo{journal}{Phys.
  Rev. Lett.} \textbf{\bibinfo{volume}{85}}, \bibinfo{pages}{1918}
  (\bibinfo{year}{2000}).

\bibitem[{\citenamefont{Emberly and Kirczenow}(2000)}]{Emberly00}
\bibinfo{author}{\bibfnamefont{E.~G.} \bibnamefont{Emberly}} \bibnamefont{and}
  \bibinfo{author}{\bibfnamefont{G.}~\bibnamefont{Kirczenow}},
  \bibinfo{journal}{Phys. Rev. B} \textbf{\bibinfo{volume}{61}},
  \bibinfo{pages}{5740} (\bibinfo{year}{2000}).

\bibitem[{\citenamefont{Boese and Schoeller}(2001)}]{Schoeller01}
\bibinfo{author}{\bibfnamefont{D.}~\bibnamefont{Boese}} \bibnamefont{and}
  \bibinfo{author}{\bibfnamefont{H.}~\bibnamefont{Schoeller}},
  \bibinfo{journal}{Europhys. Lett.} \textbf{\bibinfo{volume}{54}},
  \bibinfo{pages}{668} (\bibinfo{year}{2001}).

\bibitem[{\citenamefont{Ness et~al.}(2001)\citenamefont{Ness, Shevlin, and
  Fisher}}]{Ness01}
\bibinfo{author}{\bibfnamefont{H.}~\bibnamefont{Ness}},
  \bibinfo{author}{\bibfnamefont{S.~A.} \bibnamefont{Shevlin}},
  \bibnamefont{and} \bibinfo{author}{\bibfnamefont{A.~J.}
  \bibnamefont{Fisher}}, \bibinfo{journal}{Phys. Rev. B}
  \textbf{\bibinfo{volume}{63}}, \bibinfo{pages}{125422}
  (\bibinfo{year}{2001}).

\bibitem[{\citenamefont{Troisi et~al.}(2003)\citenamefont{Troisi, Ratner, and
  Nitzan}}]{Troisi03}
\bibinfo{author}{\bibfnamefont{A.}~\bibnamefont{Troisi}},
  \bibinfo{author}{\bibfnamefont{M.~A.} \bibnamefont{Ratner}},
  \bibnamefont{and} \bibinfo{author}{\bibfnamefont{A.}~\bibnamefont{Nitzan}},
  \bibinfo{journal}{J. Chem. Phys.} \textbf{\bibinfo{volume}{118}},
  \bibinfo{pages}{6072} (\bibinfo{year}{2003}).

\bibitem[{\citenamefont{Pecchia and {Di Carlo}}(2004)}]{Pecchia04}
\bibinfo{author}{\bibfnamefont{A.}~\bibnamefont{Pecchia}} \bibnamefont{and}
  \bibinfo{author}{\bibfnamefont{A.}~\bibnamefont{{Di Carlo}}},
  \bibinfo{journal}{Nano Lett.} \textbf{\bibinfo{volume}{4}},
  \bibinfo{pages}{2109} (\bibinfo{year}{2004}).

\bibitem[{\citenamefont{Chen et~al.}(2005)\citenamefont{Chen, Zwolak, and {Di
  Ventra}}}]{Chen05}
\bibinfo{author}{\bibfnamefont{Y.}~\bibnamefont{Chen}},
  \bibinfo{author}{\bibfnamefont{M.}~\bibnamefont{Zwolak}}, \bibnamefont{and}
  \bibinfo{author}{\bibfnamefont{M.}~\bibnamefont{{Di Ventra}}},
  \bibinfo{journal}{Nano Lett.} \textbf{\bibinfo{volume}{5}},
  \bibinfo{pages}{813} (\bibinfo{year}{2005}).

\bibitem[{\citenamefont{May and K\"uhn}(2006)}]{May06}
\bibinfo{author}{\bibfnamefont{V.}~\bibnamefont{May}} \bibnamefont{and}
  \bibinfo{author}{\bibfnamefont{O.}~\bibnamefont{K\"uhn}},
  \bibinfo{journal}{Chem. Phys. Lett.} \textbf{\bibinfo{volume}{420}},
  \bibinfo{pages}{192} (\bibinfo{year}{2006}).

\bibitem[{\citenamefont{Lehmann et~al.}(2004)\citenamefont{Lehmann, Kohler,
  May, and {H\"anggi}}}]{Lehmann04}
\bibinfo{author}{\bibfnamefont{J.}~\bibnamefont{Lehmann}},
  \bibinfo{author}{\bibfnamefont{S.}~\bibnamefont{Kohler}},
  \bibinfo{author}{\bibfnamefont{V.}~\bibnamefont{May}}, \bibnamefont{and}
  \bibinfo{author}{\bibfnamefont{P.}~\bibnamefont{{H\"anggi}}},
  \bibinfo{journal}{J. Chem. Phys.} \textbf{\bibinfo{volume}{121}},
  \bibinfo{pages}{2278} (\bibinfo{year}{2004}).

\bibitem[{\citenamefont{Koch and von Oppen}(2005)}]{Koch05}
\bibinfo{author}{\bibfnamefont{J.}~\bibnamefont{Koch}} \bibnamefont{and}
  \bibinfo{author}{\bibfnamefont{F.}~\bibnamefont{von Oppen}},
  \bibinfo{journal}{Phys. Rev. Lett.} \textbf{\bibinfo{volume}{94}},
  \bibinfo{pages}{206804} (\bibinfo{year}{2005}).

\bibitem[{\citenamefont{Wegewijs and Nowack}(2005)}]{Wegewijs05}
\bibinfo{author}{\bibfnamefont{M.~R.} \bibnamefont{Wegewijs}} \bibnamefont{and}
  \bibinfo{author}{\bibfnamefont{K.~C.} \bibnamefont{Nowack}},
  \bibinfo{journal}{New J. Phys.} \textbf{\bibinfo{volume}{7}},
  \bibinfo{pages}{239} (\bibinfo{year}{2005}).

\bibitem[{\citenamefont{Jiang et~al.}(2006)\citenamefont{Jiang, Kula, and
  Luo}}]{Jiang06}
\bibinfo{author}{\bibfnamefont{J.}~\bibnamefont{Jiang}},
  \bibinfo{author}{\bibfnamefont{M.}~\bibnamefont{Kula}}, \bibnamefont{and}
  \bibinfo{author}{\bibfnamefont{Y.}~\bibnamefont{Luo}}, \bibinfo{journal}{J.
  Chem. Phys.} \textbf{\bibinfo{volume}{124}}, \bibinfo{pages}{34708}
  (\bibinfo{year}{2006}).

\bibitem[{\citenamefont{Ryndyk et~al.}(2006)\citenamefont{Ryndyk, Hartung, and
  Cuniberti}}]{Ryndyk06}
\bibinfo{author}{\bibfnamefont{D.~A.} \bibnamefont{Ryndyk}},
  \bibinfo{author}{\bibfnamefont{M.}~\bibnamefont{Hartung}}, \bibnamefont{and}
  \bibinfo{author}{\bibfnamefont{G.}~\bibnamefont{Cuniberti}},
  \bibinfo{journal}{Phys. Rev. B} \textbf{\bibinfo{volume}{73}},
  \bibinfo{pages}{045420} (\bibinfo{year}{2006}).

\bibitem[{\citenamefont{Galperin et~al.}(2006)\citenamefont{Galperin, Ratner,
  and Nitzan}}]{Galperin06}
\bibinfo{author}{\bibfnamefont{M.}~\bibnamefont{Galperin}},
  \bibinfo{author}{\bibfnamefont{A.}~\bibnamefont{Nitzan}},
  \bibnamefont{and} \bibinfo{author}{\bibfnamefont{M.~A.} \bibnamefont{Ratner}},
  \bibinfo{journal}{Phys. Rev. B} \textbf{\bibinfo{volume}{73}},
  \bibinfo{pages}{045314} (\bibinfo{year}{2006}).

\bibitem[{\citenamefont{Gagliardi et~al.}(2007)\citenamefont{Gagliardi,
  Solomon, Pecchia, Frauenheim, {Di Carlo}, Reimers, and Hush}}]{Gagliardi07}
\bibinfo{author}{\bibfnamefont{A.}~\bibnamefont{Gagliardi}},
  \bibinfo{author}{\bibfnamefont{G.~C.} \bibnamefont{Solomon}},
  \bibinfo{author}{\bibfnamefont{A.}~\bibnamefont{Pecchia}},
  \bibinfo{author}{\bibfnamefont{T.}~\bibnamefont{Frauenheim}},
  \bibinfo{author}{\bibfnamefont{A.}~\bibnamefont{{Di Carlo}}},
  \bibinfo{author}{\bibfnamefont{N.~S.} \bibnamefont{Hush}},
  \bibnamefont{and} \bibinfo{author}{\bibfnamefont{J.~R.} \bibnamefont{Reimers}},
  \bibinfo{journal}{Phys. Rev. B} \textbf{\bibinfo{volume}{75}},
  \bibinfo{pages}{174306} (\bibinfo{year}{2007}).

\bibitem[{\citenamefont{Sergueev et~al.}(2007)\citenamefont{Sergueev, Demkov,
  and Guo}}]{Sergueev07}
\bibinfo{author}{\bibfnamefont{N.}~\bibnamefont{Sergueev}},
  \bibinfo{author}{\bibfnamefont{A.~A.} \bibnamefont{Demkov}},
  \bibnamefont{and} \bibinfo{author}{\bibfnamefont{H.}~\bibnamefont{Guo}},
  \bibinfo{journal}{Phys. Rev. B} \textbf{\bibinfo{volume}{75}},
  \bibinfo{pages}{233418} (\bibinfo{year}{2007}).

\bibitem[{\citenamefont{Frederiksen et~al.}(2007)\citenamefont{Frederiksen,
  Paulsson, Brandbyge, and Jauho}}]{Frederiksen07}
\bibinfo{author}{\bibfnamefont{T.}~\bibnamefont{Frederiksen}},
  \bibinfo{author}{\bibfnamefont{M.}~\bibnamefont{Paulsson}},
  \bibinfo{author}{\bibfnamefont{M.}~\bibnamefont{Brandbyge}},
  \bibnamefont{and} \bibinfo{author}{\bibfnamefont{A.-P.} \bibnamefont{Jauho}},
  \bibinfo{journal}{Phys. Rev. B} \textbf{\bibinfo{volume}{75}},
  \bibinfo{pages}{205413} (\bibinfo{year}{2007}).

\bibitem[{\citenamefont{Caspary-Toroker and Peskin}(2007)}]{Toroker07}
\bibinfo{author}{\bibfnamefont{M.}~\bibnamefont{Caspary-Toroker}}
  \bibnamefont{and} \bibinfo{author}{\bibfnamefont{U.}~\bibnamefont{Peskin}},
  \bibinfo{journal}{J. Chem. Phys.} \textbf{\bibinfo{volume}{127}},
  \bibinfo{pages}{154706} (\bibinfo{year}{2007}).

\bibitem[{not({\natexlab{a}})}]{note_Benesch08b}
\bibinfo{note}{C. Benesch, M. Cizek, J. Klimes, M. Thoss, and W. Domcke,
  arXiv:0712.3690 (2008).}

\bibitem[{\citenamefont{Cizek et~al.}(2004)\citenamefont{Cizek, Thoss, and
  Domcke}}]{Cizek04}
\bibinfo{author}{\bibfnamefont{M.}~\bibnamefont{Cizek}},
  \bibinfo{author}{\bibfnamefont{M.}~\bibnamefont{Thoss}}, \bibnamefont{and}
  \bibinfo{author}{\bibfnamefont{W.}~\bibnamefont{Domcke}},
  \bibinfo{journal}{Phys. Rev. B} \textbf{\bibinfo{volume}{70}},
  \bibinfo{pages}{125406} (\bibinfo{year}{2004}).

\bibitem[{not({\natexlab{b}})}]{note1}
\bibinfo{note}{For a more detailed discussion of the Hamiltonian, see Ref.\
  \onlinecite{note_Benesch08b}.}

\bibitem[{\citenamefont{Weiss}(1999)}]{Weiss99}
\bibinfo{author}{\bibfnamefont{U.}~\bibnamefont{Weiss}},
  \emph{\bibinfo{title}{Quantum Dissipative Systems}}
  (\bibinfo{publisher}{World Scientific}, \bibinfo{address}{Singapore},
  \bibinfo{year}{1999}), \bibinfo{edition}{2nd} ed.

\bibitem[{\citenamefont{Mahan}(1981)}]{Mahan81}
\bibinfo{author}{\bibfnamefont{G.~D.} \bibnamefont{Mahan}},
  \emph{\bibinfo{title}{Many-Particle Physics}} (\bibinfo{publisher}{Plenum
  Press}, \bibinfo{year}{1981}).

\bibitem[{\citenamefont{Lang and Firsov}(1963)}]{Lang63}
\bibinfo{author}{\bibfnamefont{I.~G.} \bibnamefont{Lang}} \bibnamefont{and}
  \bibinfo{author}{\bibfnamefont{Y.~A.} \bibnamefont{Firsov}},
  \bibinfo{journal}{Sov. Phys. JETP} \textbf{\bibinfo{volume}{16}},
  \bibinfo{pages}{1301} (\bibinfo{year}{1963}).

\bibitem[{\citenamefont{Keldysh}(1965)}]{Keldysh65}
\bibinfo{author}{\bibfnamefont{L.~V.} \bibnamefont{Keldysh}},
  \bibinfo{journal}{Sov. Phys. JETP} \textbf{\bibinfo{volume}{20}},
  \bibinfo{pages}{1018} (\bibinfo{year}{1965}).

\bibitem[{\citenamefont{Haug and Jauho}(1996)}]{Haug98}
\bibinfo{author}{\bibfnamefont{H.}~\bibnamefont{Haug}} \bibnamefont{and}
  \bibinfo{author}{\bibfnamefont{A.-P.} \bibnamefont{Jauho}},
  \emph{\bibinfo{title}{Quantum Kinetics in Transport and Optics of
  Semiconductors}} (\bibinfo{publisher}{Springer}, \bibinfo{address}{Berlin},
  \bibinfo{year}{1996}).

\bibitem[{\citenamefont{Langreth}(1976)}]{Langreth76}
\bibinfo{author}{\bibfnamefont{D.~C.} \bibnamefont{Langreth}},
  \emph{\bibinfo{title}{Linear and Nonlinear Electron Transport in Solids}}
  (\bibinfo{publisher}{Plenum Press}, \bibinfo{address}{New York},
  \bibinfo{year}{1976}).

\bibitem[{\citenamefont{Meir and Wingreen}(1992)}]{Meir92}
\bibinfo{author}{\bibfnamefont{Y.}~\bibnamefont{Meir}} \bibnamefont{and}
  \bibinfo{author}{\bibfnamefont{N.~S.} \bibnamefont{Wingreen}},
  \bibinfo{journal}{Phys. Rev. Lett.} \textbf{\bibinfo{volume}{68}},
  \bibinfo{pages}{2512} (\bibinfo{year}{1992}).

\bibitem[{\citenamefont{Jauho et~al.}(1994)\citenamefont{Jauho, Wingreen, and
  Meir}}]{Jauho94}
\bibinfo{author}{\bibfnamefont{A.-P.} \bibnamefont{Jauho}},
  \bibinfo{author}{\bibfnamefont{N.~S.} \bibnamefont{Wingreen}},
  \bibnamefont{and} \bibinfo{author}{\bibfnamefont{Y.}~\bibnamefont{Meir}},
  \bibinfo{journal}{Phys. Rev. B} \textbf{\bibinfo{volume}{50}},
  \bibinfo{pages}{5528} (\bibinfo{year}{1994}).

\bibitem[{\citenamefont{Mitra et~al.}(2004)\citenamefont{Mitra, Aleiner, and
  Millis}}]{Mitra04}
\bibinfo{author}{\bibfnamefont{A.}~\bibnamefont{Mitra}},
  \bibinfo{author}{\bibfnamefont{I.}~\bibnamefont{Aleiner}}, \bibnamefont{and}
  \bibinfo{author}{\bibfnamefont{A.~J.} \bibnamefont{Millis}},
  \bibinfo{journal}{Phys. Rev. B} \textbf{\bibinfo{volume}{69}},
  \bibinfo{pages}{245302} (\bibinfo{year}{2004}).

\bibitem[{\citenamefont{Benesch et~al.}(2006)\citenamefont{Benesch, Cizek,
  Thoss, and Domcke}}]{Benesch06}
\bibinfo{author}{\bibfnamefont{C.}~\bibnamefont{Benesch}},
  \bibinfo{author}{\bibfnamefont{M.}~\bibnamefont{Cizek}},
  \bibinfo{author}{\bibfnamefont{M.}~\bibnamefont{Thoss}}, \bibnamefont{and}
  \bibinfo{author}{\bibfnamefont{W.}~\bibnamefont{Domcke}},
  \bibinfo{journal}{Chem. Phys. Lett.} \textbf{\bibinfo{volume}{430}},
  \bibinfo{pages}{355} (\bibinfo{year}{2006}).

\end{thebibliography}

\end{document}